\documentstyle[aps]{revtex}

\input epsf

\begin{document}

\title{
Interface localisation--delocalisation transition in a symmetric polymer blend:\\
a finite--size scaling Monte Carlo study
} 

\author{
M.\ M\"{u}ller\footnote{Email address: \tt{Marcus.Mueller@uni-mainz.de}} and K.\ Binder
\\
{\small Institut f{\"u}r Physik, WA 331, Johannes Gutenberg Universit{\"a}t}
\\
{\small D-55099 Mainz, Germany}
}
\date{August 21, 2000 submitted to Phys.Rev.E}
\maketitle

\begin{abstract}
Using extensive Monte Carlo simulations we study the phase diagram of a symmetric binary (AB) polymer blend confined into
a thin film as a function of the film thickness $D$. The monomer--wall interactions are short ranged and antisymmetric, 
i.e, the left wall attracts the $A$--component of the mixture with the same strength as the right wall the $B$--component,
and give rise to a first order wetting transition in a semi--infinite geometry. The phase diagram and the crossover between different critical
behaviors is explored. For large film thicknesses we find a first order
interface localisation/delocalisation transition and the phase diagram comprises two critical points, which are the finite
film width analogies of the prewetting critical point. Using finite size scaling techniques we locate these critical points
and present evidence of 2D Ising critical behavior. When we reduce the film width the two critical points approach the symmetry
axis $\phi=1/2$ of the phase diagram and for $D \approx 2 R_g$ we encounter a tricritical point. For even smaller film thickness
the interface localisation/delocalisation transition is second order and we find a single critical point at $\phi=1/2$.

Measuring the probability distribution of the interface position we determine the effective interaction between the wall and
the interface. This effective interface potential depends on the lateral system size even away from the critical points. Its
system size dependence stems from the large but finite correlation length of capillary waves. This finding gives direct evidence 
for a renormalization of the interface potential by capillary waves in the framework of a microscopic model.
\end{abstract}

\section{ Introduction. }
Confining a binary mixture one can profoundly alter its miscibility behavior.\cite{EREV,PREV,DIETRICH,DEGENNES,POLYMER} If a mixture is confined into a 
quasi one--dimensional (e.g., cylindrical) pore no true phase transition occurs, unlike the prediction of the mean 
field theory. Fluctuations destroy long--range order and only a pronounced maximum of the susceptibility remains 
in the vicinity of the  unmixing transition in the bulk. In a two--dimensional system (e.g., a slit--like pore or a film) 
with identical surfaces a true phase transition occurs (capillary condensation) and the shift of the critical point away 
from its bulk value has been much investigated.\cite{NAKANISHI} The confinement changes the universality class of the 
transition from 3D Ising critical behavior in the bulk to  2D Ising critical behavior in the film. The latter 
manifests itself in much flatter binodals in a film close to the unmixing transition than in the bulk. No such
change of the critical exponents is observed in mean field theory.

The phase behavior of symmetric mixtures in a thin film with antisymmetric surface interactions has attracted abiding interest 
recently.\cite{BROCHARD,PE,SWIFT,BINDER,TROUBLE,GRAVITY}
The right surface attracts one species with exactly the same strength as the opposite surface attracts the other species.
In contrast to capillary condensation, the phase transition does not occur close to the unmixing transition in the bulk, but rather in the vicinity
of the wetting transition. Close to the unmixing transition in the bulk, enrichment layers at the surfaces are gradually built up
and an interface is stabilized in the middle of the film. In this ``soft--mode'' phase the system is laterally homogenous
-- no spontaneous breaking of the symmetry occurs. If the wetting transition of the semi--infinite system is of second order
one encounters a second order localisation--delocalisation transition slightly below the wetting transition temperature.
The system phase separates laterally into regions where the interface is located close to one surface (localized state). The order parameter,
i.e., the distance between the interface and the center of the film, grows continously. This prediction of
phenomenological theories has been corroborated by detailed simulation studies\cite{BINDER,NEW,ANDREAS} and it is also in accord with 
experimental findings.\cite{KLEIN,SF}

If the wetting transition is of first order and the thickness of the film not too small, mean field calculations\cite{MSCF1,MSCF2} predict the 
occurance of two critical points which correspond to the prewetting critical point of the semi--infinite system. Unlike the 
wetting transition,\cite{NAKANISHI} the prewetting transition can produce a critical (singular) behavior in a thin film, because only the 
lateral correlation length diverges at the prewetting critical point; the thickness of the enrichment layers at the surfaces remains finite.
The mean field treatment invokes approximations and it cannot be expected to capture the subtle interplay between 2D Ising fluctuations at the 
critical points, ``bulk''--like composition fluctuations, and interface fluctuations typical for the wetting transition.\cite{NEW} Consequently, 
a detailed test of the mean field predictions via Monte Carlo simulations is certainly warranted and elucidates the role of fluctuations.
Using Monte Carlo simulations of the Ising model Ferrenberg {\em et al.} studied the interface localisation--delocalisation transition 
also for the case that the wetting transition of the semi--infinite system is of first order.\cite{B2} The simulation study was centered on the 
dependence on the film thickness, which is a convenient parameter to be varied in experiments.  However, the study was restricted to the
coexistence between strictly symmetric phases and many questions remained open.  

The general features of the phase behavior are shared by
all binary mixtures. Here, we present large scale Monte Carlo simulations aiming at investigating the phase behavior of a symmetric binary polymer 
blend confined between antisymmetric walls. Computationally, simulations of a polymer blend\cite{MREV} are much more demanding than studying simple
fluids (e.g., the Ising model), but recent mean field calculations made detailed predictions for the phase behavior of confined polymer 
mixtures\cite{MSCF1,MSCF2} and serve as guidance for choosing the model parameters in the simulations. Simulating polymer blends, we can, at least 
in principle, control the importance of fluctuations by varying 
the degree of interdigitation, i.e., the chain length.\cite{MSCF2,MREV} The mean field theory is expected to become accurate
in the limit of infinite interdigitation. In a binary polymer blend the wetting transition occurs at much lower temperatures than the critical temperature
of the unmixing transition in bulk.\cite{WET} Hence, ``bulk''--like composition fluctuations are not important in the vicinity of the wetting transition 
temperature and we can isolate the effect of interface fluctuation. Moreover, these systems are also suitable candidates to examine the phase behavior 
experimentally.  Indeed, one of the first studies of the ``soft--mode'' phase has employed a binary polymer blend.\cite{KLEIN}

Our paper is broadly arranged as follows: First, we present a phenomenological description of the phase behavior in a film with antisymmetric 
short range surface interactions. Using a standard model for the effective interface potential we calculate the phase behavior in mean field approximation,
discuss the regime of validity of the mean field approach, and consider the crossover between the different critical behaviors. Second, we briefly describe 
our coarse grained lattice model for a binary polymer mixture. Then, we present our Monte Carlo results: We obtain the phase diagram for film thicknesses 
ranging from $D=1.1 R_g$ to $7 R_g$, where $R_g$ denotes the radius of gyration of the polymer chains, investigate the critical behavior and present evidence
that interface fluctuations renormalize the effective interface potential.
We close with a comparison of the phase diagram to the behavior of the bulk and of films with  symmetric boundary conditions.

\section{Background.}
Rather than describing the configuration of the system by the detailed composition profile across the film, much qualitative insight into
the thermodynamics can be deduced from the effective interface potential. Below the bulk--critical temperature enrichment layers of the prefered
components form at the surfaces and stabilize an $AB$ interface which runs parallel to the walls. The effective interface potential $g_{\rm wall}(l)$
describes the
free energy per unit area as a function of the distance $l$ between this $AB$ interface and a wall. In the case of short range forces between the
monomers and the walls, the interface profile becomes distorted in the vicinity of the walls and this gives rise to an interaction which
decays exponentially as a function of the distance l between the $AB$ interface and a single wall:
\begin{equation}
g_{\rm wall}(l) = a\exp(-\lambda l) - b\exp(-2\lambda l) + c \exp(-3\lambda l)
\label{eqn:sw}
\end{equation}
This expression retains only the lowest powers of $\exp(-\lambda l)$, which are necessary to bring about a first order wetting transition of the semi--infinite
system. The coefficient $a$ is explicitely temperature dependent, while the temperature dependence of $b$ and $c$ is neglected. $c>0$ is assumed throughout the discussion.
All coefficients are of the same magnitude as the interfacial tension $\sigma$ between the coexisting bulk phases. For polymer blends this quantity
scales with chain length $N$ and monomer number density $\rho$ like $\sqrt{\bar N}/R_g^2$.
$\bar N = (\rho R_g^3/N)^2$ measures the degree of interdigitation. $1/\lambda$ denotes the spatial range of the interactions and it is of the 
order $R_g$.  $b<0$ gives rise to a second order
wetting transition at $a=0$; and $b=0$ to a tricritical transition. For $b>0$ one encounters a first order wetting transition at $a_{\rm wet}=b^2/4c$ where
the thickness of the enrichment layer jumps discontinuously
from $l_-=1/\lambda \ln (2c/b)$ to a macroscopic value.\cite{MSCF2} The wetting spinodals take the values $a>0$ (from the wet phase) and $a< b^2/3c$ (from the
non--wet phase). The concomitant prewetting line terminates at the prewetting critical point $a_{\rm pwc} =16 a_{\rm wet}/9$ and 
$l_{\rm pwc} = 1/\lambda \ln (9c/2b)$. 

We approximate the effective interface potential in a film to be the linear superposition of the interactions originating 
at each wall and analyze the behavior. SCF calculations\cite{MSCF2} lend support to this approximation. The interface potential 
in a film of thickness $D$ takes the form:
\begin{eqnarray}
g(l) & = & g_{\rm wall}(l) + g_{\rm wall}(D-l) -2 g_{\rm wall}(D/2) \nonumber \\
     & = &   2 a \exp(-\lambda D/2) \left(\cosh \left( \lambda [l-D/2] \right)-1\right)
       - 2 b \exp(-\lambda D) \left(\cosh \left( 2 \lambda [l-D/2] \right)-1\right) \nonumber \\
     &&  + 2 c \exp(-3 \lambda D/2) \left(\cosh \left( 3 \lambda [l-D/2] \right)-1\right)
\end{eqnarray}
In general, the phase boundaries depend on the variables $a/c,b/c$ and $\lambda D$.
If we proceeded as in the Ref\cite{NEW} by expanding the $\cosh$ in powers of $[l-D/2]$, the further analysis 
would be rather cumbersome.  A more transparent procedure employs the variable
\begin{equation}
\tilde{m}^2 = 2 \exp(-\lambda D/2) \left(\cosh \left( \lambda [l-D/2] \right)-1\right) = 
     \left( \exp(-\lambda D/4) \lambda[l-D/2] \right)^2 + \mbox {higher orders of}\quad [l-D/2]
\end{equation}
to rewrite the interface potential in the form
\begin{equation}
g(l) = c \left[ \tilde{m}^2 (\tilde{m}^2-r)^2 + t\tilde{m}^2 \right] \qquad \mbox{with} \qquad 
r=\frac{b-6c\exp(-\lambda D/2)}{2c} \qquad \mbox{and} \qquad 
t = \frac{a-a_{\rm wet}-b\exp(-\lambda D/2)}{c}
\label{eqn:landau}
\end{equation}
The qualitative form of the effective interface potential has been inferred previously on the basis of a Landau expansion,\cite{MSCF2} here it is derived
explicitely from the standard form of the interface potential (\ref{eqn:sw}) for a first order wetting transition in the semi--infinite system. Negative values
of $r$ correspond to  second order localization--delocalization transitions, $r=0$ to a tricritical one, and positive values of $r$ give rise to first order
transitions. $t$ measures the distance from the tricritical transition temperature (for $r\leq 0$) and denotes the triple temperature in the case of a first
order interface localization--delocalization transition (cf.\ below). For $r\leq 0$ the phase boundaries depend only on the two parameter combinations $r$ 
and $t$. In these variables the limit $\lambda D \to \infty$ is particularly transparent: $cr \to b/2$, $ct \to a-a_{\rm wet}$ and $\tilde{m} \to 
\exp(-\lambda l)$.

\subsection{$r \leq 0$: second order and tricritical interface localization--delocalization transition}
A second order interface localization--delocalization transition (i.e. $r<0$) will occur either if the wetting transition is second order 
(i.e., $b<0$) or if the wetting transition is first order but the film thickness $D$ small enough to comply with $0<b< 6c\exp(-\lambda D/2)$. 
This behavior is in accord with previous findings\cite{SWIFT,B2,MSCF2} and we shall corroborate this further by our present simulations.
Since the coexisting phases are symmetric with respect to exchanging $l$ and $D-l$, phase coexistence occurs at $\Delta \mu_{\rm coex}\equiv 0$ or 
$\partial g/\partial l = (\partial g/\partial \tilde{m}) ({\rm d}\tilde{m}/{\rm d}l) = 0$. From this condition we obtain for the binodals:
\begin{equation}
\tilde{m}^2 = \frac{2|r|}{3} \left(\sqrt{1+\frac{3}{4r^2}\Delta t}-1 \right) \to \left\{\begin{array}{ll}
                                                                              \Delta t/4|r| & \mbox{for} \quad \Delta t \ll r^2 \qquad ({\rm 2DMF}) \\
									      \sqrt{\Delta t/3} &  \mbox{for} \quad \Delta t \gg r^2 \qquad ({\rm 2DTMF}) 
                                                                              \end{array}
                                                                       \right.
\label{eqn:5}
\end{equation}
The critical temperature is given by $t_c=-r^2$ and $\Delta t = t_c-t$ denotes the distance from the critical temperature at fixed $r$.
For $r<0$ the binodals at the critical point open with the mean field exponent $\beta_{\rm 2DMF}=1/2$. This corresponds to mean field critical behavior (2DMF)
of a system with a single scalar order parameter, i.e., $m = [l/D-1/2]$. At larger distance the order parameter grows like 
$m \sim (\Delta t)^{\beta_{\rm 2DTMF}}$
with $\beta_{\rm 2DTMF}=1/4$. The latter exponent is characteristic for the mean field behavior at a tricritical point (2DTMF). The crossover between
mean field critical and tricritical behavior occurs around $|\Delta t_{\rm cross}| \sim r^2$. As we decrease the magnitude of $r \to 0$ we approach the 
tricritical point and the regime where mean field critical behavior is observable shrinks. At the tricritical point only the tricritical regime (2DTMF) 
exists, i.e., $\Delta t_{\rm cross}=0$, and the binodals take the particularly simple form $\tilde{m}=(\Delta t/3)^{1/4}$. The crossover in the binodals 
for $r=-0.4$ is illustrated in the inset of Fig.\ref{fig:cross}({\bf a}).

Of course, the above considerations neglect fluctuations and the behavior close to the transition is governed by Ising critical exponents and two dimensional
tricritical exponents, respectively. The crossover between Ising critical behavior (2DI) and tricritical behavior (2DT) occurs at
$|\Delta t_{\rm cross}| \sim r^{1/\phi_{\rm cross}}$, where the crossover critical exponent is not $1/2$ (as for the crossover between the mean field
regimes) but rather $4/9$.\cite{TRI1,TRI2,TRI3} Following Ref\cite{NEW} we calculate the critical amplitudes and estimate the location of the crossover between mean field
critical behavior and the region where fluctuations dominate the qualitative behavior. For small values of the order parameter $m=[l/D-1/2]$ we approximate
$m \approx \tilde{m} \exp(\lambda D/4)/(\lambda D)$ and obtain for the mean field critical amplitudes: 
\begin{equation}
\hat B_{\rm 2DMF}=\frac{\exp(\lambda D/4)}{2\sqrt{|r|}\lambda D} \qquad  \mbox{and} \qquad \hat B_{\rm 2DTMF}=\frac{\exp(\lambda D/4)}{3^{1/4}\lambda D}
\end{equation}
The susceptibility of the order parameter above the critical temperature is related to the inverse curvature of the interface potential in the middle of the film
$1/\chi D^2 = \frac{\partial^2 g}{\partial l^2}_{|l=D/2}$. Using Eq(\ref{eqn:landau}) we obtain for critical and tricritical mean field transitions:
\begin{equation}
\chi = \frac{1}{2 c (\lambda D)^2}\exp(\lambda D/2) \Delta t^{-1} \qquad \mbox{and} \qquad \hat C^+_{\rm MF} = \frac{1}{2 c(\lambda D)^2}\exp(\lambda D/2) \qquad \gamma_{\rm MF}=1
\end{equation}
The ratio $\hat C^+_{\rm MF}/\hat C^-_{\rm MF}$ of the critical amplitudes above and below the critical point is universal and takes the mean field value 2
at the critical point and 4 at the tricritical point.
At the transition the correlation length $\xi_\|$ diverges. This lateral length is associated with fluctuations of the local interface position, i.e.,
capillary waves. In mean field approximation the parallel correlation length takes the form:
\begin{equation}
\xi_\| = \left(\frac{1}{\sigma}\frac{\partial^2 g}{\partial l^2}\right)^{-1/2} = \sqrt{\sigma D^2 \chi} \qquad \mbox{hence} 
\qquad \hat \xi^+_{\rm MF} = \frac{\sqrt{\sigma}}{\lambda \sqrt{2c}} \exp(\lambda D/4) \qquad \gamma_{\rm MF}=1/2
\end{equation}
and $\hat \xi^+_{\rm MF}/\hat \xi^-_{\rm MF}=\sqrt{2}$ and $2$, respectively.

Knowing the critical amplitudes we can estimate the importance of fluctuations via the Ginzburg criterium:\cite{GINZBURG} As it is well known, 
mean field theory is self--consistent if the fluctuations of the order parameter in a volume of linear dimension $\xi_\|$ are small in comparison 
to the mean value of the order parameter. For our quasi--two--dimensional system $(d=2)$ we obtain:
\begin{equation}
\frac{\chi}{\xi_\|^d} \stackrel{!}{\ll} m^2  \qquad \Rightarrow  \qquad 
\left(\frac{c^{1-2/d}\lambda^2}{\sigma} \right)^{d/2} \exp(-d \lambda D/4) \stackrel{!}{\ll} \left\{\begin{array}{lll}
              \frac{1}{|r|} \Delta t ^{(4-d)/2} & \mbox{for} \qquad r<0 & \mbox{second order}\\
              \Delta t ^{(3-d)/2} & \mbox{for} \qquad $r=0$ & \mbox{tricritical}
                                                                                                            \end{array}
                                                                                                     \right.
\end{equation}
This result is as expected:
For our quasi--to--dimensional system we obtain for a second order interface localization--delocalization 
transition $\Delta t \ll Gi_{\rm 2DI} \sim |r| \exp(-\lambda D/2) / \sqrt{\bar N}$ in accord with Ref\cite{NEW}, while we obtain
$\Delta t \ll Gi_{\rm 2DT} \sim \exp(-\lambda D)/\bar N$ upon approaching the tricritical point.
For bulk ($d=3$) tricritical phenomena Landau theory is marginally correct.

Combining the above results we find the following behavior upon approaching the critical temperature:
Far away from the tricritical point, i.e., $r \gg \exp(-\lambda D/2)/\sqrt{\bar N}$ we find mean field tricritical behavior (2DTMF) for $\Delta t \gg r^2$,
mean field critical behavior (2DMD) for $ r^2 \ll \Delta t \ll |r| \exp(-\lambda D/2)$, and finally two dimensional Ising critical behavior (2DI)
for $|r| \exp(-\lambda D/2) \gg \Delta t$. Closer to the tricritical point, i.e.  $r \ll \exp(-\lambda D/2)/\sqrt{\bar N}$, we find  mean field tricritical 
behavior (2DTMF) for $\Delta t \gg \exp(-\lambda D/4)$,  two dimensional tricritical behavior (2DT) for $\exp(-\lambda D/4) \ll \Delta t \ll C r^{1/\phi_{\rm cross}}$,
and Ising critical behavior (2DI) for $C r^{1/\phi_{\rm cross}} \gg \Delta t$. The prefactor $C$ must be chose such that all crossover lines
( 2DI $\leftrightarrow$ 2DT, 2DT $\leftrightarrow$ 2DTMF, 2DTMF $\leftrightarrow$ 2DMF, and 2DMF $\leftrightarrow$ 2DI) intersect in a common
point. This yields $C= (\bar N \exp(\lambda D))^{-1+1/2\phi_{\rm cross}}$. Of course, the term ``crossover line'' is not meant as a sharp division
between different behaviors, but should be understood rather as a center of a smooth crossover region. Likewise, the above constant $C$ may involve 
a constant of order unity which has been suppressed for simplicity.
The two different sequences can be clearly distinguished in the
Monte Carlo simulations, because the probability distribution of the order parameter exhibits a three peak structure\cite{NIGEL} only close to the
tricritical point (2DT). We shall use this property to distinguish between the two different sequences in our MC simulations. The anticipated behavior 
is summarized in Fig.\ref{fig:cross}({\bf a}).

In the Monte Carlo simulation this rich crossover scenario is further complicated by finite size rounding. The Monte Carlo results are subjected 
to pronounced finite size effects whenever the correlation length becomes of the order of the lateral system size. In the mean field regime the
correlation length scales like $\xi_\| \sim R_g \exp(\lambda D/4) \Delta t^{-1/2}$. Knowing the Ginzburg number for the crossover from 2DMF
to 2DI behavior we estimate the correlation length in the Ising critical regime:\cite{NEW}
\begin{equation}
\xi_\| \sim R_g \exp(\lambda D/4) \Delta t^{-1/2} \tilde f(\Delta t /Gi_{\rm 2DI}) \to \left\{\begin{array}{ll}
                  R_g \exp(\lambda D/4) \Delta t^{-1/2} & \mbox{for} \qquad \Delta t \gg Gi_{\rm 2DI} \\
		  R_g |r|^{1/2}\bar N^{-1/4} \Delta t^{-1} & \mbox{for} \qquad \Delta t \ll Gi_{\rm 2DI} 
\end{array}\right.
\end{equation}
where we have assumed that the scaling function $\tilde f$ assumes a power law behavior for small and large arguments and we have used the value
$\nu_{\rm 2DI}=1$ appropriate for the divergence of the correlation length in the 2DI regime. 
Similarly, we determine the correlation length in the 2DT regime:
\begin{equation}
\xi_\| \sim R_g \exp(\lambda D/4) \Delta t^{-1/2} \tilde f(\Delta t /Gi_{\rm 2DT}) \to \left\{\begin{array}{ll}
                  R_g \exp(\lambda D/4) \Delta t^{-1/2} & \mbox{for} \qquad \Delta t \gg Gi_{\rm 2DT} \\
		  R_g \exp(\lambda D(3/4-\nu_{\rm tri})) \bar N^{1/2-\nu_{\rm tri}} \Delta t^{-\nu_{\rm tri}} & \mbox{for} \qquad \Delta t \ll Gi_{\rm 2DT} 
\end{array}\right.
\end{equation}
$\nu_{\rm tri}=5/9$ denotes the exponent of the correlation length in the 2DT universality class.\cite{TRI1,TRI2,TRI3} The correlation lengths
at the various crossovers are compiled in Tab.1. The largest correlation length occurs at the crossover from 2DT to 2DI behavior
\begin{equation}
\xi_\|^{2DT \leftrightarrow 2DI} \sim R_g \exp(\lambda D(3/4-\nu_{\rm tri}/2\phi_{\rm cross})) \bar N^{1/2-\nu_{\rm tri}/2\phi_{\rm cross}} |r|^{-\nu_{\rm tri}/\phi_{\rm cross}}
\label{eqn:xl}
\end{equation}
In order to observe the true Ising critical behavior for negative values of r, the system size $L$ has to exceed this correlation length. 
In the vicinity of the tricritical point (i.e., for small negative values of $r$) this requirement is very difficult to be met in computer 
simulations.

\subsection{$r>0$: first order interface localization--delocalization transition}
For positive values of $r$ the interface potential exhibits a three valley structure. The three minima at $\tilde{m}=\pm\sqrt{r}$ and $\tilde{m}=0$ have equal free energy
at $t=0$. This corresponds to the triple point. At lower temperatures an $A$--rich phase coexist with a $B$--rich phase, and since the two phases are 
symmetrical the coexistence occurs at $\Delta \mu_{\rm coex}=0$. The binodals below the triple point take the form:
\begin{equation}
\tilde{m} = \pm \sqrt{\frac{2r}{3}+\sqrt{\frac{r^2}{9}-\frac{t}{3}}} \qquad \mbox{for} \qquad t<0,\qquad r>0
\label{eqn:tt}
\end{equation}

Above the triple temperature $t>0$ there are 2 two phase coexistence regions symmetrically located around $\tilde{m}=0$. 
These phase coexistences terminate in two critical points. Since the coexisting phases correspond to a thick and a thin enrichment layer of the
prefered phase at each wall, there is no symmetry between the coexisting phases, and the exchange potential $\Delta \mu_{\rm coex}$ at coexistence
differs from zero. Unfortunately, the phase boundaries for $t>0$ and $r>0$ depend not only on $r$ and $t$ but also on $\lambda D$ explicitely, and
we have determined them numerically. The dependence of the critical temperature $t_c$ on $r$ for several values of
$\lambda D$ is presented in Fig.\ref{fig:cross}({\bf b}). The coexistence curve for $b/c=4.44$ and various values of $\lambda D$ are presented in
the inset of Fig.\ref{fig:cross}({\bf b}). As the film thickness is decreased the critical temperature decreases and the critical points
move closer to the symmetry axis of the phase diagram. They are determined by the condition:
\begin{equation}
\frac{\partial^2g}{\partial l^2} = \frac{\partial^3g}{\partial l^3} = 0 \qquad \mbox{at} \qquad t=t_c  \qquad \mbox{and} \qquad \tilde{m}=\tilde{m}_c
\end{equation}

In two limiting cases a simple behavior emerges: 

(i) If $|\lambda(l-D/2)| \ll 1$ we can replace the derivative w.r.t.\ $l$ by derivatives w.r.t.\ $\tilde{m}$ and obtain:\cite{MSCF2}
$t_c = 7r^2/5$ and $\tilde{m}_c=\pm\sqrt{2r/5}$. This approximation holds for $r \ll \exp(-\lambda D/2)$. Expanding $g$ in powers of
$\delta \tilde{m} = \tilde{m}-\tilde{m}_c$ we obtain (omitting an irrelevant term linear in $\delta \tilde{m}$):
\begin{equation}
g(\tilde{m}) \approx c(t_c-t)\;\delta \tilde{m}^2 +4rc\;\delta \tilde{m}^4 +\frac{7c}{3} \;\delta \tilde{m}^5 + c\;\delta \tilde{m}^6
\label{eqn:15}
\end{equation}
This allows us to calculate the binodals in the vicinity of the critical points, the susceptibility and parallel correlation length.
The presence of a 5th order term in $\delta \tilde{m}$ in the expansion (\ref{eqn:15}) is a manifestation of the fact that the phase
boundaries of the prewetting transitions are not symmetric around $\tilde{m}_c$. This lack of symmetry is also evident from the
numerical results in Fig.\ref{fig:cross}({\bf b}) (insert).
The critical amplitudes scale in the same way with $r$, $\bar N$, and $\lambda D$ as for $r<0$. In particular we find for the crossover
between 2DMF behavior and 2DI behavior $Gi_{\rm 2DI} \sim |r|\exp(-\lambda D/2)/\sqrt{\bar N}$.

(ii) In the limit of large film thickness $\lambda D \to \infty$, the critical point tends towards the prewetting critical point 
at $t_c = t_{\rm pwc}=7r^2/9$. In this limit confinement effects are negligible and the coexistence curves in the vicinity of
the critical points corresponds to the prewetting lines at the corresponding surfaces. We expect the same critical behavior as
at the prewetting critical point. In this case the Ginzburg number does not depend on the film thickness. For $\lambda D \to \infty$ we employ
the interface potential at a single wall, and find for the validity of the mean field description:
\begin{equation}
\frac{a-a_{\rm pwc}}{a_{\rm pwc}} \gg \frac{\lambda^2}{\sigma} \sim \bar N^{-1/2} \sim Gi_{\rm 2DI} \qquad \mbox{for} \qquad r>0 \qquad \lambda D \to \infty
\end{equation}

\section{ The bond fluctuation model and simulation technique.}
Modeling polymeric composites from the chemical details of the macromolecular repeat units to the morphology of the phase separated blend within a single model
is not feasible today even with state-of-the-art supercomputers.  Yet, there is ample evidence that by a careful choice of simulation and analysis techniques,
coarse grained models of flexible polymers  -- like the bond fluctuation model\cite{MREV,BFM} -- provide useful insights into the universal polymeric features. 
In the framework of the bond fluctuation
model each effective monomer blocks a cube of 8 neighboring sites from further occupancy on a simple cubic lattice in three dimensions. Effective monomers
are connected by bond vectors of length $2,\sqrt{5},\sqrt{6},3$, or $\sqrt{10}$ in units of the lattice spacing. The bond vectors are chosen such that
the excluded volume condition guarantees that chains do not cross during their motion.\cite{HPD} Each effective
bond represents a group of $n\approx3-5$ subsequent $C-C$-bonds along the backbone of the chain.\cite{MAP} Hence, the chain length
$N=32$ employed in the present simulations corresponds to a degree of polymerization of $100-150$ in a real polymer. If we increased the chain
length $N$, the mean field theories would yield a better description of the equilibrium thermodynamics (self--consistent field theory is believed to be 
quantitatively accurate in the limit $N \to \infty$) but the length scale of the ordering phenomena would be larger. Hence, our choice of $N$ is a compromise 
determined by the computational resources.  The statistical segment length $b$ in the relation for the radius of gyration $R_g=b\sqrt{N/6}$ is 
$b=3.05$ (i.e., $R_g\approx 7$ for $N=32$).  

We study thin films of geometry $ L \times L \times D$. Periodic boundary conditions are applied in the two lateral directions, while there are hard impenetrable 
walls at $z=0$ and $z=D+1$ modeling a film of thickness $D$. The average number density in the film is $\rho_0=1/16$, i.e., half of the lattice sites are occupied by
corners of monomers. This density corresponds to a melt or concentrated solution. The density profile of occupied lattice sites, normalized by the bulk value, is 
presented in Fig.\ref{fig:DENS} for film thickness $D=24$ and $48$. For this choice of temperature and monomer--wall interaction an interface is stabilized in the 
center of the film.  Due to the extended shape of the monomers and the compressibility of the fluid there are packing effects at the walls.\cite{WET} Overall
the walls are repulsive and the monomer density is slightly reduced in the boundary region. The spatial extension of this region is independent of the film thickness.
Moreover, the density is reduced at the center of the interface as to reduce the energetically unfavorable contacts between unlike species.\cite{MBO} 
Both effects are not incorporated into the mean field calculations\cite{MSCF1,MSCF2} 
and cause the density in the ``bulk''--like region of the film to be slightly larger for thinner films than for thicker ones. In the following we employ the density 
of occupied lattice sites in the layers $5 \leq z \leq 8$ as a measure of the density of the film. For large $D$ the data are compatible with a behavior of the form
$\rho = \rho_0(1+0.85/D)$. The film thickness ranges from  $D=12 \approx 1.7 R_g$ to $D=48 \approx 7 R_g$ and we vary the lateral extension over a wide range 
$48 \leq L \leq 264$ to
analyze finite size effects. In the two layers nearest to the walls, monomers experience a monomer--wall interaction. An $A$--monomer is attracted by the left 
wall and repelled
by the right wall, the interaction between $B$--monomers and the walls is exactly opposite. Each monomer--wall interaction changes the energy by an amount
$\epsilon_w=0.16$ in units of $k_BT$. For these parameters the wetting transition and the phase diagram of a blend confined between symmetric walls has been investigated 
previously.\cite{WET} 

Binary interactions between monomers are catered for by a short ranged square well potential $\epsilon \equiv -\epsilon_{AA}=-\epsilon_{BB}=\epsilon_{AB} \equiv 1/k_BT$ which is 
extended up to a distance $\sqrt{6}$. The phase separation is brought about by the repulsion between the unlike species. The Flory-Huggins parameter is 
$\chi=2z_{\rm eff}\epsilon$ where $z_{\rm eff}\approx 2.65$ denotes the effective coordination number in the bulk\cite{M0,MREV} at $\rho_0=1/16$. 
For $\epsilon_w=0.16$ previous simulations find a strong first order wetting transition at $T_{\rm wet}= 1/\epsilon_{\rm wet}=14.1(7)$.\cite{WET}
This value corresponds to $\chi N \approx 12$, which is well inside the strong segregation limit.

The polymer conformations are updated via a combination of random monomer displacements and slithering snake--like movements. The latter relax the chain 
conformations about a factor of $N$ faster than the local displacements.\cite{M0} We work in the semi--grandcanonical ensemble,\cite{SARIBAN} i.e., we control the temperature
$T \equiv 1/\epsilon$ and the exchange potential $\Delta \mu$ between the two species, and the concentration fluctuates. This semi--grandcanonical ensemble
is realized in the Monte Carlo simulations via switching the polymer identity $A \rightleftharpoons B$ at fixed chain conformation. The different Monte Carlo moves are applied in the
ratio: slithering snake : local displacements : semi--grandcanonical identity switches = 12:4:1. During production runs, we record all 150 slithering snake 
steps the composition, the energy, and the surface energy and obtained the joint probability distribution in form of a histogram.
We use the semi--grandcanonical identity switches in junction with a reweighting scheme,\cite{MBO,REWEIGHT}
i.e., we add to the Hamiltonian of the system a reweighting function ${\cal H}_{\rm rw} =  {\cal H}_{\rm orig}+W(\phi)$, which depends only on the overall
composition $\phi= n_{\rm poly}^A/(n_{\rm poly}^A+n_{\rm poly}^B)$. $n_{\rm poly}^A$ and $n_{\rm poly}^B$ denote the number of $A$ and $B$ polymers in the simulation cell, respectively. The choice
$W(\phi) \approx -\ln P(\phi)$ , where $P(\phi)$ denotes the probability distribution of the composition in the semi--grandcanonical ensemble, encourages 
the system to explore configurations 
in which both phases coexist in the simulation cell. Otherwise these configurations would be severely suppressed due to the free energy cost of interfaces. In the framework 
of this reweighting scheme the system ``tunnels'' often from one phase to the other and this allows us to locate the phase coexistence accurately and measure 
the free energy of the mixture as a function of the composition $\phi$.  Use of histogram extrapolation technique\cite{HISTO} permits histograms obtained at one set of model 
parameters to be reweighted to yield estimates appropriate to another set of model parameters. We employ this analysis technique to obtain estimates for the 
reweighting function $W(\phi)$.

\section{ Results. }
Firstly, we locate the critical points of the phase diagrams. For very small film thickness we find a second order localisation--delocalisation transition
even though the wetting transition is of first order. Swift {\em et al} have predicted this behavior in the framework of a square gradient theory\cite{SWIFT} and such
a behavior is also born out in our self--consistent field calculations for polymer blends\cite{MSCF1,MSCF2} and simulations of the Ising model.\cite{B2} Upon increasing the film thickness
we encounter a nearly tricritical transition. A truly tricritical transition cannot be achieved by tuning the film thickness only, because of the discreteness 
of the lattice, but it could be brought about by varying the monomer--wall interaction. In an experiment using real materials, of course, the film thickness
can be varied continously, and a truly tricritical transition is in principle always accessible.
For even larger film thickness the interface localisation--delocalisation 
is first order and we find two critical points at $\phi \neq 1/2$.

Secondly, we locate the triple line for the two largest values of the film thicknesses and discuss how capillary waves lead to a strong dependence of the effective 
interface potential on the lateral system size.

Thirdly, we detail our results on the thickness dependence of the phase diagram and relate our findings to the binodals of the bulk and the mixture confined 
into a film with symmetric boundaries.

\subsection{ Critical points. }

\subsubsection{$D=8 \approx 1.1 R_g$ and $D=12\approx 1.7 R_g$: second order interface localisation--delocalisation transition}
For film thicknesses which are comparable to the radius of gyration of the molecules, the effective interface potential originating from the two surfaces
strongly interfere. This might change the order of the interface localisation--delocalisation transition from first to second. In this case, a single critical 
point occurs on the symmetry axis $\phi=1/2$ of the phase diagram. The transition is thought to belong to the 2D Ising universality class. In 
Fig.\ref{fig:2DI} ({\bf a}) we present the probability distribution of the composition for various inverse temperatures $\epsilon$, film thickness
$D=8$ and lateral film extension $L=80$. Upon increasing the monomer--monomer interaction $\epsilon$ the probability distribution $P(\phi)$ changes 
from single--peaked to bimodal, which indicates that a phase transition occurs in this temperature range. No signature of the trimodal distribution
occurs and, hence, we conclude that the system is far away from the tricritical point, i.e., $|r| > \exp(-\lambda D/2)/\sqrt{\bar N}$. In this case,
we expect a crossover from 2DMF to 2DI behavior.

Along the coexistence curve $\Delta \mu=0$ and its extension to higher temperatures we use the cumulant intersection method to locate the critical point.\cite{FSS}
In the vicinity of the critical point the probability distribution of the order parameter $m=\phi-\phi_{\rm coex}=\phi-1/2$ scales to leading order like:\cite{FSS}
\begin{equation}
P(m,L,t) \sim L^{\beta/\nu}{ P}^\star(L^{\beta/\nu}m,L^{-1/\nu}t)
\end{equation}
where $t=(\epsilon_c-\epsilon)/\epsilon_c$ denotes the distance from the critical point along the coexistence curve and $\beta$ and $\nu$ are the 
critical exponents of the order parameter and the correlation length. $P^\star$ is characteristic of the universality class and has been obtained 
from simulations of the Ising model\cite{BRUCEWILDING} at the critical temperature $t=0$.  Cumulants of the form $\langle m^2\rangle/\langle |m|\rangle^2$ 
are expected to exhibit a common intersection point for different system sizes $L$ at the critical temperature.\cite{FSS} The value of the cumulant
at the intersection point is universal. Our simulation data are presented in panel ({\bf b}) and exhibit some corrections to scaling due to the crossover 
2DMF to 2DI behavior. Similar corrections were observed in simulations of a second order interface localization/delocalization transition in the Ising 
model.\cite{NEW} From the intersection points of neighboring system sizes and from the intersection of the cumulant with the universal value of the Ising
model we estimate the critical temperature to be $\epsilon_c=0.0520(5)$.

In the inset of Fig.\ref{fig:2DI}({\bf b}) we show the probability distribution normalized to unit variance and norm at our estimate of the
critical temperature $\epsilon_c=0.052$, and compare the distribution to universal scaling curve of the 2D Ising universality class. The
probability distributions for the smaller system sizes are slightly broader than the universal scaling curve, but the deviations decrease as
we increase the system size. 

The simulation data for $D=12$ are presented in Fig.\ref{fig:2DI}({\bf c}) and ({\bf d}). As we lower the temperature the probability distribution of the
composition for $L=48$ changes from single peaked to bimodal.
At intermediate values of $\epsilon$, however, a three--peak structure is clearly discernable. This is characteristic of the 2DT regime and 
indicates the vicinity of the tricritical point. In the phenomenological considerations this regime occurs only for $|r| < \exp(-\lambda D/2)/\sqrt{\bar N}$. 
We note that the distribution for that small lateral system sizes resembles at no value of $\epsilon$ the universal shape of the order parameter distribution 
of the 2D Ising model. We conclude that the finite size rounding for this lateral system size sets in before we observe the crossover from  2DT
to 2DI behavior, i.e., the correlation length $\xi_\|^{2DT \leftrightarrow 2DI}$ in Eq (\ref{eqn:xl}) exceeds the lateral systems size $L$. For such 
small lateral extensions the universal properties of the transition are completely masked.  Larger system sizes and a careful finite size scaling analysis is 
indispensable to determine the type of transition and accurately locate the transition temperature.

The temperature dependence of the cumulant is presented in Fig.\ref{fig:2DI}({\bf d}). There is no unique intersection point and the value of the
cumulants at the crossing is larger than the universal value of the cumulant of the Ising class. This behavior indicates pronounced corrections to 
scaling due to the crossover from 2DT behavior away from the critical point to 2DI behavior at the critical point. From the intersection 
points of neighboring system sizes and from the intersection of the cumulant with the universal value of the 2D Ising model 
we estimate the critical temperature to be $\epsilon_c=0.0589(10)$.

The inset of panel ({\bf d}) compares the distribution of the order parameter at our estimate of the critical temperature  and the Ising scaling function.
As we increase the lateral system size the ``third'' peak in the distribution vanishes and $P(\phi)$ gradually approaches the universal scaling curve. This 
indicates that our largest system sizes exceed the correlation length at the crossover from 2DT to 2DI behavior. The comparison of $P(\phi)$ 
with the universal scaling curve for several system sizes accurately locates the critical point and gives evidence that the transition belongs to the 2D Ising 
universality class. 

For $D \leq 12$ we find a single interface localisation--delocalisation transition of second order at $\phi=1/2$.

\subsubsection{\bf $D=14 \approx 2 R_g$: tricritical interface localisation--delocalisation transition}
The three--peak structure in the probability distribution for $D=12$ and small lateral extensions $L$ has indicated the vicinity of the tricritical interface
localisation--delocalisation transition. Increasing the film thickness we need larger and larger lateral extensions to observe the 2DI behavior
as $\xi_\|^{2DT \leftrightarrow 2DI}$ diverges.
Right at the tricritical point the distribution of the composition is expected to exhibit a three--peak structure for all lateral system sizes and the distribution,
when scaled to unit variance and norm, coincides with a universal scaling function. Wilding and Nielaba\cite{NIGEL} have obtained this scaling function via simulations at the
tricritical point of the spin-1 Blume--Capel model\cite{BC} in two dimensions. Assuming that the tricritical interface localisation--delocalisation transition belongs to the same 
universality class, we vary the film thickness $D$ and the interaction strength $\epsilon$ as to match the probability distribution of the composition onto the
predetermined scaling function of the tricritical universality class. This strategy largely facilitates the search of the tricritical interface 
localisation--delocalisation transition. Fig.\ref{fig:TRI}({\bf a}) displays the probability distribution of the composition for film thicknesses ranging
from $D=12$ to $D=18$ and the universal scaling curve. The temperature was adjusted for each film thickness such that the relative heights of the central and outer 
peaks correspond to the ratio of the universal scaling curve. For small $D<D_{\rm tri}$ the ``valley'' between the peaks is too shallow and for $D>D_{\rm tri}$ the 
probability between the peaks is too small. For $D \gg D_{\rm tri}$ this situation corresponds to the triple point (cf.\ below) and the probability of finding a system
between the peaks is suppressed by the free energy cost of the interface between the phases with composition close to $0$ or $1$ and the ``soft--mode'' phase with
composition $\phi=1/2$. As we increase the film thickness the temperature at which the ratio between the peak height equals $1.2$ shifts
towards lower temperatures and approaches the wetting transition temperature from above. 

Panel ({\bf a}) of Fig.\ref{fig:TRI} suggests that the tricritical transition occurs close to the film thickness
$D=14$. This is further corroborated in Fig.\ref{fig:TRI}({\bf b}), where we show the distribution function at $\epsilon=0.06151$ for various system sizes. Within the
statistical accuracy of our data the distribution functions for the larger systems sizes collapse well onto the universal scaling curve. For smaller systems
the outer peaks are slightly sharper and centered at smaller values of the order parameter.
Of course, no perfect data collapse can be expected because we can tune the film thickness only in units of the lattice spacing.
In view of the statistical accuracy and possible systematic corrections to scaling, however, we did not attempt to vary the monomer--wall interaction 
$\epsilon_w$ as to achieve a better collapse. For $D=14$ the system is very close to the tricritical transition.

\subsubsection{ $D=24\approx 3.5 R_g$ and $D=48 \approx 7R_g$: critical points for $\phi \neq 1/2$}
Though the system is strictly symmetric the critical points for larger film thickness ($D>D_{\rm tri}$) do not occur at $\phi=1/2$ but rather there are two critical 
points at critical compositions $\phi_c$ and $1-\phi_c$. These critical points are the finite film thickness analogs of the prewetting critical points, which
occur in the limit $D \to \infty$.\cite{MSCF1}
Below the critical temperature the phase diagram comprises two miscibility gaps. The coexisting phases 
correspond to surfaces with a thin and a thick enrichment layer of the preferred component. Due to the missing symmetry between the coexisting phases the coexistence
value of the chemical potential $\Delta \mu_{\rm coex}$ differs from zero. We determine $\Delta \mu_{\rm coex}$ via the equal weight rule,\cite{EWR} i.e., we adjust $\Delta \mu$ such 
that
\begin{equation}
\int_0^{\phi^\star} {\rm d}\phi \; P(\phi) \stackrel{!}{=} \int_{\phi^\star}^1 {\rm d}\phi \; P(\phi) 
\qquad \mbox{and} \qquad
\phi^\star = \int_0^1 {\rm d}\phi \; P(\phi) \phi
\end{equation}

Along this coexistence curve and its finite--size extension to higher temperatures we use the cumulant intersection to locate the critical temperature.
This is shown in Fig.\ref{fig:TCD48}({\bf a}) for the film thickness $D=24$. For the system sizes accessible in the simulations the 
intersection points between cumulants of neighboring systems sizes systematically shift to lower temperatures and the value of the cumulant at the intersection 
point gradually approaches the value of the 2D Ising universality class from above. The latter is indicated in the figure by the horizontal line. From these data we estimate
the critical parameters to be $\epsilon_c=0.061(1)$, and $\phi_c=0.18(2)$ and $\phi_c=0.82(2)$ respectively. This corresponds to a critical thickness 
$l_c=D \phi_c = 0.62 R_g$ of the enrichment layer. A similar procedure has been employed to locate the critical temperature in the film of thickness $D=48$. The temperature 
and system size dependence of the cumulants are displayed in Fig.\ref{fig:TCD48}({\bf c}). From this we extract the estimate $\epsilon_c=0.0625(10)$ for the critical 
temperature and $\phi_c=0.09(2)$ and $\phi_c=0.91(2)$ for the critical compositions. This value corresponds to a distance between then wall and the interface of 
$l_c=0.63 R_g$. Since increasing the film thickness from $3.5 R_g$ to $7R_g$ does not change $T_c$ or $l_c$ substantially, we are in the regime $\lambda D \gg 1$
and the critical behavior is characteristic of the prewetting critical point in the semi--infinite system.

The behavior of the cumulants and the very gradual approach of the probability distribution towards the Ising curve indicate pronounced corrections to scaling. 
For the simulation of the bulk phase diagram\cite{HPD2} a nice cumulant intersection has been obtained with system sizes in the range $24^3$ to $56^3$. In the present
study we employ systems with about an order of magnitude more polymers and obtain no clear intersection of the cumulants! There are three reasons for strong corrections to 
the leading 2D Ising scaling behavior:
(i)  The aspect ratio $D/L$ of our simulation cell is always finite. Truly two--dimensional behavior can only be observed for vanishing aspect ratio, and our data
     might fall into the broad crossover region between three--dimensional critical behavior and the two--dimensional critical behavior. Such a crossover has been 
     studied in our polymer model for neutral walls\cite{YAN} and walls, which attract both the same species (i.e., capillary condensation).\cite{WET}  However, we note that unlike these situations
     there is no three--dimensional critical point in the vicinity for antisymmetric boundary conditions. The temperature of the unmixing transition in the bulk 
     is a factor 4 higher than the critical point in a thin film. Since the critical point in a thin film is related to the prewetting transition of the 
     semi--infinite system, i.e., a transition with no three dimensional analogy, we expect the corrections to be qualitatively different from the case 
     of neutral or symmetric boundaries.
(ii) Unlike the situation for small film thickness $D=12$ the probability distribution of the order parameter is asymmetric, because the critical point does not 
     lay on the symmetry axis of the phase diagram. This missing symmetry between the two phases gives rise to field--mixing effects,\cite{BRUCEWILDING} which manifest themselves in
     corrections of relative order $L^{-(1-\alpha-\beta)/\nu}$. These corrections are antisymmetric to leading order and, hence, are not expected to influence even 
     moments (like the cumulants) of the order parameter distribution profoundly. The effects are, however, detectable in the order parameter distribution 
     which we present in Fig.\ref{fig:TCD48}({\bf b}) and ({\bf d}). The distribution functions at our estimate of the critical temperature clearly lack
     symmetry and approach very gradually the symmetric scaling 
     curve of the 2D Ising universality class.
(iii) Additionally, there are corrections to scaling by non--singular background terms.  One source of (non--critical) composition fluctuations are ``bulk--like'' 
     fluctuations in the $A$--rich and $B$--rich domains. In a bulk system, i.e., with periodic boundary conditions in all directions, the susceptibility is rather 
     small.  At $\epsilon=0.065$ it takes the value $\chi_T^{\rm bulk}=V \langle \Delta \phi^2 \rangle =  0.047$ with $\Delta \phi = \phi - \langle \phi \rangle$. 
     In a system of size $96 \times 96 \times 24$ this 
     susceptibility corresponds to composition fluctuations of the order $\sqrt{\langle \Delta \phi^2 \rangle} \sim 5 \cdot 10^{-4}$. Therefore, we believe that 
     ``bulk--like'' composition fluctuations are not the major source of background terms. However, we cannot rule out that the presence of an $AB$ interface gives 
     rise to enhanced composition fluctuations.
     Another source of corrections to scaling stems from the fluctuations in the average interface position itself. Since the effective interaction between the 
     interface and the wall is rather weak, they give rise to a finite but large susceptibility away from the critical point. 
     We have estimated the susceptibility from the curvature of $\ln P(\phi)$ close to the triple point (i.e., $T \approx 0.9 T_c$), and we have
     obtained values of the order $\chi_T \sim 3 \cdot 10^2$ (and a smaller value is obtained if the interface is close to a wall.) For the same 
     system size as above, this yields  composition fluctuations of the order  $\sqrt{\langle \Delta \phi^2 \rangle} \sim 0.04$ (a value which should be compared
     to $\phi_c(D=24)=0.18(2)$). This observation partially 
     rationalizes why the peak in the probability distribution of the composition close to $\phi=1/2$ is always broader than the peak which corresponds to the 
     phase in which the interface is close to the wall.  As we approach the critical temperature composition fluctuations grow. At the critical point 
     the typical composition fluctuations are of the order $\sqrt{\Delta \phi^2} \sim \sqrt{L^{\gamma/\nu-d}} \sim L^{-1/8}$, where we have used the critical 
     exponents for 
     the susceptibility $\gamma = 7/4$ and the correlation length $\nu=1$ appropriate for the 2D Ising universality class. Hence, for small system sizes typical
     fluctuations yield compositions which differ substantially from the critical composition; only for very large sizes the composition fluctuates in the
     vicinity of the critical value. Moreover, the critical density is much displaced from the symmetry axis $\phi=1/2$ and typical fluctuations in a finite system 
     are cut-off by the constraint $0 < \phi$ or $\phi<1$. Therefore, the susceptibility of a small system is reduced compared to the value expected from the 
     leading scaling behavior. 
     This observation is in accord with our Monte Carlo data, and a similar reasoning has been used by Bruce and Wilding\cite{BWE} in discussing background terms to the
     specific heat and the concomitant corrections to scaling in the energy distribution.


\subsection{ The triple point. }
For the largest two film thicknesses $D=24$ and $D=48$ the interface localisation---delocalisation transition is first order and the concomitant
two miscibility gaps join in a triple point. At this temperature an $A$-rich phase, a $B$--rich phase and a phase where the interface is located in the
middle of the film ($\phi=1/2$) coexist. The coexisting phases correspond to three peaks in the distribution of the composition. Upon increasing the lateral 
system size the peak positions do not shift (as opposed to the behavior at the tricritical point), the peaks become more pronounced and configurations with 
intermediate compositions are more and more suppressed, because of the presence of interfaces between the coexisting phases.

The composition of the system and the average interface position are related via  $l=\phi D$ (integral criterium), where we assume that
the coexisting bulk phases are almost pure, i.e., $\phi_{\rm coex}^{\rm bulk}\approx 0$ or $1$. From the probability distribution we then calculate 
the effective interface potential $g(l)$:
\begin{equation}
g(l) = -\frac{k_BT}{L^2} \ln P(\phi=l/D)
\end{equation}
In principle, not only fluctuations of the interface position $ \langle \Delta l^2 \rangle $ but also ``bulk''--like fluctuations contribute
to composition fluctuations $\langle \Delta \phi^2 \rangle \approx \frac{1}{D^2} \langle \Delta l^2 \rangle + \frac{\chi_T^{\rm bulk}}{L^2D}$.
Since the wetting transition in a binary polymer blend occurs far below the critical point of the bulk, the bulk susceptibility is very small, and
the latter contribution can be neglected.

The dependence of the free energy per unit area on the position of the interface is a key ingredient into the theory of wetting.\cite{CAHN,MICHAEL,DIETRICH,LIPOWSKI,BREZIN,OMEGA1,OMEGA2} 
The interface interacts with the boundaries and the (bare) interface potential exhibits three minima. These correspond to the three coexisting phases. 
In the two phases with $\phi$ close to $0$ and $1$, the interface is localized close to the wall, the interaction between the wall and the interface is rather
strong, and the effective interface potential possesses a deep minimum. In the ``soft--mode'' phase the interface is only weakly bound to the center 
of the film and the minimum is much broader. In Fig.\ref{fig:REND} we present the effective interface potentials for film thicknesses $D=24$ ({\bf a}) and
$D=48$ ({\bf b}) and various lateral system sizes in the vicinity of the triple temperature. The three minima are clearly visible, however, the shape of 
the interface potential and the value of the minima depend on the lateral system size $L$. Moreover, the minima which correspond to the localized states
broaden and (slightly) shift to larger distances between wall and interface upon increasing $L$ (cf.\ inset).

Fluctuations of the local interface position, i.e., capillary waves, lead to a renormalization of the effective interface potential $g(l)$ and
cause the dependence of $g(l)$ on the lateral system size, which we observed in a microscopic model of a polymer mixture. Describing the configuration of the 
system only via the local position $l(x,y)$ of the interface (sharp kink approximation) we write the coarse grained free energy in form of the capillary wave 
Hamiltonian:\cite{LIPOWSKI,OMEGA1,IH}
\begin{equation}
{\cal H}[l] = \int {\rm d}^2x\;  \left\{ \frac{\sigma}{2} (\nabla l)^2  +  g(l)  \right\}
\label{eqn:eih}
\end{equation}
where $\sigma$ approaches the $AB$ interface tension between the coexisting bulk phases for large separations between the wall and the interface. 
An increase of $\sigma$ at smaller distances $l$ as revealed by previous MC simulations is neglected.\cite{WET} In the vicinity of a minimum of $g(l)$ we may 
approximate the interface potential by a parabola.
\begin{equation}
g(\delta l) = {\rm const}+\frac{1}{2} \sigma k_\|^2 \delta l^2
\end{equation}
$\delta l$ denotes the deviation of the local interface position from the position where the $g(l)$ attains its minimum. $\xi_\|=2\pi/k_\|$ is
the parallel correlation length of interface fluctuations. For lateral distances much smaller than $\xi_\|$ the fluctuations of the local
interface position are hardly perturbed by the interaction between the interface and the wall; the interface behaves like a free interface.
For lateral distances which exceed $\xi_\|$ capillary waves are strongly suppressed. $\xi_\|$ is larger for the minimum of $g(l)$ in the center
of the film than for the minima, in which the interface is localized at a wall. From the curvature of the effective interface potential $g(l)$ 
for film thickness $D=24$ we estimate $k_1=\sqrt{ \frac{{\rm d}^2g}{{\rm d}\phi^2}/\sigma D^2}=0.26$ and $k_2=0.031$, where we have used the bulk 
value $\sigma=0.0382$ for the interfacial tension at $\epsilon=0.068$. For the thicker film we obtain $k_1=0.3$, but the curvature in the middle of the film
could not be accurately estimated. The value is of the order $k_2 \sim {\cal O}(0.005)$, and we expect this value to decrease exponentially
with the film thickness. Hence, this fluctuation effect is the stronger the larger the film thickness. For the system sizes employed in the MC 
simulations $k_\| L$ is of order unity.

In our Monte Carlo simulations the finite lateral system size $L$ acts as an additional cut--off for the spectrum of interface fluctuations\cite{ANDREAS} and 
upon increasing $L$ we extend the spectrum of interface fluctuations. Allowing for interface fluctuations we decrease the free energy of the system.
Therefore, we expect the free energy density of the system to decrease when we increase the lateral system size, and we expect the effect to be the
stronger the larger $\xi_\|$. Therefore, the free energy of the ``soft--mode'' phase becomes smaller compared to the free energy of the phase, where 
the interface is located close to a wall when we increase $L$. This effect is clearly observed in the MC simulations. To be more quantitative, we 
consider a system where the laterally averaged interface position is at the minimum of $g(l)$, and we expand the deviation $\delta l(x,y)$ from the 
minimum in a Fourier series
\begin{equation}
\delta l(x,y) = \sum_{n,m=0}^\infty \left\{
                              a_{nm} \cos (q_n x) \cos(q_m y)
                            + b_{nm} \cos (q_n x) \sin(q_m y)
                            + c_{nm} \sin (q_n x) \cos(q_m y)
                            + d_{nm} \sin (q_n x) \sin(q_m y)
                              \right\}
\label{eqn:exp}
\end{equation}
with $q_n = 2\pi n/L$. The coefficients $a_{00}=b_{00}=c_{00}=d_{00}=b_{0m}=c_{0m}=d_{0m}=d_{n0}$ vanish identically, all other coefficients
can take any real value. Using this expansion (\ref{eqn:exp}) and the effective interface Hamiltonian (\ref{eqn:eih}), we calculate the average
size of fluctuations
\begin{equation}
\langle a^2_{nm} \rangle = \frac{4}{\sigma (Lk_\|)^2} \left[1 + \left(\frac{2\pi}{Lk_\|} \right)^2 (n^2+m^2) \right]^{-1}
\end{equation}
and the free energy
\begin{eqnarray}
\frac{F}{k_BTL^2} &=& - \frac{1}{L^2} \ln \int {\cal D}[l] \exp\left(\frac{-{\cal H}[l]}{k_BT}\right) \nonumber \\
 &=& {\rm const} + \frac{2}{L^2} \sum_{nm=0}^\infty \eta_{nm} \ln \left\{\frac{\sigma}{k_BT} \left[ k_\|^2 + \left(\frac{2\pi}{L}\right)^2(n^2+m^2) \right] \right\} 
\end{eqnarray}
where the factor $\eta_{nm}$ takes the values $\eta_{00}=0$, $\eta_{n0}=\eta_{0m}=1/2$ and $\eta_{nm}=1$ for $n \neq0$ and $m \neq 0$
in order to account for the restriction on the coefficients $a,b,c,d$. The additive constant is independent of the wavevector cut-off $k_\|$. 
The dependence of the free energy on the system size is dominated by the small $q$ behavior. In this regime the discrete nature of the wavevector 
space matters and, hence, we do not replace the sum over $q$ by integrals. Using the measured values of the wavevector cut-offs we calculate the
lateral system size dependence of the free energy difference between the ``soft--mode''phase and the delocalized state. The results are compared
to the MC data in Fig.\ref{fig:D24}. Good agreement is found for large $L$, whereas there are deviations for smaller $L$. For small $L$ the 
amplitude of the fluctuations becomes large and the a parabolic interface potential is no longer a good approximation -- especially for the localized state
where the interface is located very close to the walls.
We have used histogram extrapolation to adjust the temperature such that the difference $\Delta g(l)=g_2-g_1$ of the minima
vanishes. This corresponds to the equal height criterium for the triple point. The equal weight condition, which we have applied to determine the binodals
close to the critical points, would require: $\Delta g = \frac{1}{L^2} \ln \frac{k_1}{k_2}$. Both conditions agree, of course, when we extrapolate our results 
to $L \to \infty$. From this procedure we obtain the following estimates for the triple point: $1/\epsilon_{\rm triple} =14.7(4) $ and 
$\phi_{\rm triple} = 0.015,\; 0.5,\; 0.985$ for $D=24$ and $1/\epsilon_t= 14.2(4) $ and $\phi_{\rm triple} = 0.0066,\; 0.5,\; 0.9934$ for $D=48$. 
The thickness of the microscopic enrichment layer at the wetting transition temperature is of the order $l_{\rm wet}=0.05 R_g$; a value which is consistent
with expectation for strong first order wetting transitions.

The dependence of the critical temperature and the triple temperature on the film thickness is summarized in Fig.\ref{fig:TC}.
When we increase the film thickness the critical temperature $1/\epsilon_c$ shows a non--monotonic dependence. At $D=14$ the tricritical point 
(where the critical temperature and the triple temperature merge) occurs at $\epsilon_{\rm tri}=0.0615(5)$, for film thickness
$D=24$ we find $\epsilon_c=0.0610(10)$ and at $D=48$ $\epsilon_c=0.0625(10)$. This effect is rooted in two opposing effects. On the one hand,
the self--consistent field calculations predict the Flory--Huggins parameter $\chi_c^{\rm SCF}(D)$ to decrease upon increasing the film thickness $D$ for
an incompressible fluid. This shift in temperature decreases exponentially with the film thickness. One the other hand, packing effects, which are not incorporated 
in the self--consistent field calculations, increase the density 
in the ``bulk''--like portion of the film when we decrease the film thickness. These packing effects at the walls depend strongly on the computational model, 
but qualitatively similar effects might occur in experimental systems as well. This thickness dependence of the density in the middle of the film modifies the 
relation between the depth of the square well potential and the $\chi$--parameter. This leads to a behavior of the form $\epsilon_c \sim \chi_c^{\rm SCF} / 
( 1 + 0.85/D)$, where we use the dependence of the density profile (cf.\ Fig.\ref{fig:DENS}) on the film thickness as obtained by direct measurement in the
Monte Carlo simulations. A dependence of the fluid packing structure on the density is neglected.
A similar $1/D$ correction to the difference in surface free energies between the $A$ and the $B$--rich phase has been observed in previous
simulations.\cite{WET}. 
Attempting to separate these two effects we also present $[(1+0.85/D)\epsilon_c]^{-1}$ which corresponds to the inverse Flory--Huggins 
parameter. Within the error bars the behavior of this quantity is consistent with the mean field prediction. The critical value of the inverse Flory--Huggins
parameter increases and the triple value decreases as we increase the film thickness. The latter approaches the wetting transition temperature\cite{WET} 
$T_{\rm wet}=14.1(7)$ from above.

\subsection{ The phase diagram. }
For film thickness $D=48$ we have determined the complete phase diagram. Close to the critical point we assume 2DI behavior with an exponent 
$\beta = 1/8$ for the order parameter and employ finite size scaling to estimate the critical amplitude. Outside the critical region but above the triple
temperature we have estimated the location of the binodals via the equal weight criterium in a system of size $L=64$; but no finite size analysis has been
applied. The phase diagram for a blend confined into a film with antisymmetric walls is presented in Fig.\ref{fig:PHASEN}. ({\bf a})
Confinement into a film with antisymmetric boundary conditions enlarges the one phase region up to the prewetting critical temperature. Since the wetting
transition in binary polymer blends occurs far below the unmixing critical temperature in the bulk the effect is quite pronounced. The temperature region
between the prewetting critical point and the triple point is about $11\%$ of the wetting transition temperature. This value strongly depends on the
details of the structure at the walls. The stronger the wetting transition the larger are the prewetting lines and the more extended is the region of the two
miscibility gaps. The phase diagram of the bulk and a film with symmetric walls are displayed for comparison in Fig.\ref{fig:PHASEN}. The symmetric film has the 
same thickness as the antisymmetric film and the monomer--wall interactions at one wall are identical and attract the $A$--component. While the prewetting
at the wall which prefers the $A$--component leads to a two phase region in the antisymmetric case there is only a change in curvature of the binodal detectable 
in the symmetric case.

Panel ({\bf b}) of Fig.\ref{fig:PHASEN} presents the phase diagram as a function of temperature and exchange chemical potential.
In the antisymmetric case $\Delta \mu_{\rm coex} = 0$ up to the triple temperature. There, two coexistence lines emerge which are the thin film analogies of the
prewetting lines at the two walls. Since the monomer--wall interactions are short ranged the prewetting line in the bulk and the coexistence curves in the film 
deviate from the bulk coexistence value linearly (up to logarithmic corrections).\cite{HS} They end in two critical points.
Though the system is strictly symmetric with respect to exchanging $A \rightleftharpoons B$ phase coexistence is not restricted to $\Delta \mu=0$, and the 
coexisting phases are not related by the symmetry of the Hamiltonian. The coexistence curve of the symmetric film is shown for comparison. The coexistence 
value of the chemical potential is shifted to values disfavoring the component attracted by both walls. There is a change in the temperature dependence of the 
coexistence curve close to the wetting transition temperature, but the coexistence curve stays far away from the prewetting line. If the two lines intersected 
there would also be a triple point in the symmetric case.\cite{WET,TRIPLE1} Since, the shift of the chemical potential $\Delta \mu$ is roughly proportional to the inverse film 
thickness (Kelvin equation) we expect a triple point to occur only for much larger film thicknesses. This is in accord with self--consistent field calculations.\cite{WET}
The typical distance $l$ between the interface and the wall at coexistence is of order $D/2$ in the antisymmetric case, while it is only of the order $R_g \ln D/R_g$ 
in the symmetric case. Hence, smaller film thicknesses are sufficient to study the interaction between the interface and the wall, and antisymmetric boundary
conditions are computationally more efficient to investigate the wetting behavior.

\section{ Summary and discussion. }
We have studied the phase diagram of a symmetric polymer mixture in a thin film with antisymmetric boundary conditions via large scale Monte Carlo simulations.
The walls interact with monomers via a short range potential; the one wall attracts the $A$ component and repels the $B$ component while the interaction at
the opposite wall is exactly reversed. The salient features of the phase diagram and its dependence on the film thickness as obtained by our MC simulations are 
in accord with the results of mean field theory.\cite{SWIFT,MSCF1,MSCF2} Fluctuations, which are neglected in the mean field calculations, do not modify the 
qualitative phase behavior. However, they give rise to a rich crossover behavior between Ising critical behavior, tricritical behavior and their mean field 
counterparts. This has been elucidated by phenomenological considerations and is qualitatively consistent with our simulation results.

Since the critical point of the thin binary polymer film occurs at much lower temperature than the unmixing transition in the bulk, ``bulk--like'' composition
fluctuations are only of minor importance. The dominant fluctuations of the composition of the film arise from capillary waves at the interface between the
$A$--rich and $B$--rich regions in the film. The interaction between the walls and the interface is rather small, because it is mediated via the distortion of 
the interface profiles at the walls and the strength of the interaction decreases exponentially with the distance. Hence, the interface is only very weakly bound to 
the minimum of the effective interface potential. These large
fluctuations give rise to rather pronounced corrections to scaling in our systems of limited size. However, using the cumulant intersection method\cite{FSS} and the 
matching of the order parameter distribution onto the predetermined universal scaling function,\cite{BRUCEWILDING} we give evidence for the 2D Ising universal 
character of the
critical points. The same strategy has proven computationally very convenient to locate the tricritical point as a function of the film thickness.\cite{NIGEL}
This technique allows us to locate the critical points of the confined complex fluid mixture with an accuracy of a few percent.

Interface fluctuation do not only impart 2D Ising critical behavior onto the critical points, but they are important in the whole temperature range. Monitoring
the probability distribution of the laterally averaged interface position we extract the effective interface potential $g(l)$. Its dependence on the lateral
system size yields direct evidence for the renormalization of the interface potential by interface fluctuations. Interface fluctuations lead to a broadening
of the minima in the interface potential, a shift of the minima towards the center of the film, and to a relative reduction of the free energy of the broader 
minimum. This leads to a systematic overestimation of the triple temperature by the mean field calculations.

Moreover, our simulations indicate that packing effects in thin films result in corrections of the order $1/D$ to the density of the film
or to the effective Flory--Huggins parameter. Such corrections are likely to mask completely the subtle thickness dependence of the triple temperature 
and the triple temperature predicted by the mean field calculations. For short range interactions between walls and monomers the predicted shifts
decrease exponentially with the film thickness $D$. However, power--law dependencies are expected for the case of long range (i.e., van
der Waals) interactions between walls and monomers.

The gross features of the phase diagram as well as our simulation and analysis techniques are not restricted to binary polymer fluids but generally 
apply to  binary liquid mixtures in confined geometries. Moreover, mean field calculations\cite{MSCF1} indicate that for small deviation from perfectly
antisymmetric boundary conditions a qualitatively similar phase behavior emerges. The stronger the first order wetting transitions at the boundaries,
the larger deviations from antisymmetry are permissible without alternating the topology of the phase diagram. Hence, a thin binary film on a substrate
against air/vacuum, where the substrate energetically favors one component of the mixture while the other component has an affinity to the air surface,
is an experimental realization of the boundary conditions discussed here. Our findings also imply that ultrathin enrichment layers at one surface are 
unstable in the temperature range $T_{\rm wet} < T < T_c$. Such effects have been observed experimentally\cite{EXP1} in polymeric films, although 
for a liquid--vapor transition instead of a liquid--liquid demixing. However, recent experiments have observed the wetting transition in binary polymer 
blends.\cite{EXP3,EXP2}

\subsection*{Acknowledgement}
It is a great pleasure to thank N.B.\ Wilding for stimulating discussions and for providing the universal probability distributions of the order parameter
at the 2D Ising critical and 2D tricritical point. We have benefitted also from discussions with E.V.\ Albano and A.\ De Virgiliis. Financial support by the 
DFG under grant Bi 314/17 
in the priority program ``wetting and structure formation at interfaces'' and the DAAD/PROALAR2000 as well as generous grants of computing time at the 
NIC J{\"u}lich, the HLR Stuttgart and the computing center in Mainz are gratefully acknowledged.

\newpage
\begin{table}
       \caption{
       \label{tab:1}
Compilation of the boundaries of the different regimes in the vicinity of the tricritical point and the
correlation lengths at the crossover. The latter quantity gives an estimate of the system size required
to observe the crossover in the Monte Carlo simulations.
       }
\begin{tabular}{|l|l|l|}
crossovers & $|\Delta t_{\rm cross}|$ & $\xi_{\rm cross}/R_g$ \\
\hline
2DT $\leftrightarrow$ 2DI    & $(\bar N \exp(\lambda D))^{-1+1/2\phi_{\rm cross}} r^{1/\phi_{\rm cross}}$ &
$\exp(\lambda D (3/4-\nu_{\rm tri}/2\phi_{\rm cross})) \bar N^{1/2-\nu_{\rm tri}/2\phi_{\rm cross}} r^{-\nu_{\rm tri}/\phi_{\rm cross}} $\\
2DI $\leftrightarrow$ 2DMF   & $|r| \bar N^{-1/2} \exp(-\lambda D/2)$ & 
$ |r|^{-1/2} \bar N^{1/4} \exp(\lambda D/2)$ \\
2DMF$\leftrightarrow$ 2DTMF  & $ r^2$ &
$ |r|^{-1} \exp(\lambda D/4)$ \\
2DTMF $\leftrightarrow$ 2DT  & $\bar N^{-1} \exp(-\lambda D)$ & 
$\bar N^{1/2} \exp(3\lambda D/4)$\\
\end{tabular}
\end{table}

\newpage
\begin{figure}[htbp]
    \begin{minipage}[b]{160mm}%
       \caption{
       \label{fig:cross} 
       ({\bf a}) Illustration of the different regimes for a second order and tricritical transition:
       2DTMF: mean field tricritical behavior, 
       2DMD:  mean field critical behavior,
       2DI:   two dimensional Ising critical behavior, and
       2DT:   two dimensional tricritical behavior.
       The inset shows the temperature dependence of the order parameter $\tilde{m}$ for $r=-0.4$ as
       calculated within mean field theory (see Eq.(\protect\ref{eqn:5})).
       For \protect{$t_c-t \ll 16r^2/3$} 2DMF behavior is found, while 2DTMF behavior
       is observed at larger distances from the critical point.
       ({\bf b}) Dependence of the critical temperature $t_c$ on the distance $r$ from the tricritical point.
       The curves correspond to different values of $\lambda D$ as indicated in the key. Thick lines, which bracket 
       the behavior, correspond to $t_c=7r^2/5$ (valid  for small $r$) and $t_c=7r^2/9$ (valid in the limit $\lambda 
       D \to \infty$). The inset presents the binodals at fixed strength $b=4.44$ of the wetting transition of the
       individual surface and several values of $\lambda D$ as indicated in the key. For choice of parameter 
       $b/c=4.44>3\exp(-\lambda D/2)$ (and, hence, $r>0$) there are two critical points for all values of the film thickness. 
       }
    \end{minipage}%
\end{figure}

\begin{figure}[htbp]
    \begin{minipage}[b]{160mm}%
       \caption{
       \label{fig:DENS} 
       Density of blocked lattice sites normalized by the bulk value as a function of the distance from the wall at $\epsilon=0.06$ and $\epsilon_w=0.16$
       for film thicknesses $D=24$ and $D=48$.
       Note the strong packing effects at the wall for $z \leq 5$. For these parameters an interface is stabilized at the center of the film. The position
       of the interface fluctuates in the interval $R_g\approx 7 < z < D-R_g$ (cf.\ Fig.\protect\ref{fig:REND}) The inset presents the normalized
       density averaged over the layers 5-8. This region is marked by the arrow in the main panel.
       }
    \end{minipage}%
\end{figure}

\begin{figure}[htbp]
    \begin{minipage}[b]{160mm}%
       \caption{
       \label{fig:2DI} 
       ({\bf a}) Probability distribution of the composition for system size $D=8$ and $L=80$. The inverse
       temperatures are indicated in the key.  Histogram reweighting has been applied to extrapolate the data
       along the coexistence curve. The shape of the distribution function changes from single--peaked to
       bimodal, but there is no indication of a third peak at $\phi=1/2$.
       ({\bf b}) Cumulant ratio $\langle m^2 \rangle/\langle |m| \rangle^2$ along the coexistence curve $\Delta \mu=0$
       for film thickness $D=8$ and various lateral extensions L as indicated in the key. In the finite size scaling limit 
       $L \to \infty$, $t \equiv (\epsilon-\epsilon_c)/\epsilon_c \to 0$, $Lt$ finite, the cumulant intersection should occur 
       at the value $\langle m^2 \rangle / \langle |m|\rangle^2=1.072$, highlighted by the horizontal straight line.
       Our estimate of the critical temperature $\epsilon_c=0.0520(5)$ is indicated by the double arrow. 
       The inset shows the distribution function of the order parameter -- scaled to unit norm and variance --
       at our estimate of the critical temperature and compares the MC results to the universal
       distribution of the 2D Ising universality class.
       ({\bf c}) Same as ({\bf a}) but for system size $D=12$ and $L=48$. Note that there is a broad range of $\epsilon$ where 
       the distribution has three peaks, unlike the Ising model. This indicates the vicinity of the tricritical point.
       ({\bf d}) Cumulant ratio $\langle m^2 \rangle/\langle |m| \rangle^2$ for film thickness $D=12$ and various lateral extensions L as 
       indicated in the key.  Our estimate of the critical temperature $\epsilon_c=0.0589(10)$ is indicated by the double arrow.
       The inset shows the distribution function of the order parameter at our estimate of the critical temperature and compares the 
       MC data to the universal distribution of the 2D Ising universality class.
       }
    \end{minipage}%
\end{figure}

\begin{figure}[htbp]
    \begin{minipage}[b]{160mm}%
       \caption{
       \label{fig:TRI} 
       ({\bf a})
       Probability distribution of the composition for various film thicknesses as indicated in the key.
       The lateral systems size is $L=96$. We have adjusted the interaction strength $\epsilon$
       such that the central peak is a factor $1.2$ higher than the outer peaks. In accordance with convention
       we have scaled the distributions to unit norm and variance. Circles mark the universal distribution of 
       2D tricritical transition.
       ({\bf b}) Temperature dependence of the cumulants for $D=14$ and lateral system sizes as indicated in the key.
       The horizontal line marks the cumulant value of the universal tricritical distribution.
       ({\bf c}) Probability distribution of the composition at $\epsilon_{\rm tri} = 0.06151(50)$ scaled to unit
       norm and variance. The universal 2D tricritical distribution (from Wilding and Nielaba\protect\cite{NIGEL}) is shown for comparison
       }
    \end{minipage}%
\end{figure}

\begin{figure}[htbp]
    \begin{minipage}[b]{160mm}%
       \caption{
       \label{fig:TCD48} 
       ({\bf a})
       Temperature dependence of the cumulant $\langle m^2 \rangle/\langle |m|\rangle^2$
       for $D=24$ and various system sizes as indicated in the key. The arrow marks the
       critical temperature range $\epsilon_c=0.061(1)$.
       ({\bf b})
       Probability distribution of the composition scaled to unit norm and variance
       at our estimate of the critical temperature $\epsilon=0.061$. Thin lines denote 
       the results of the Monte Carlo simulations. Histogram reweighting has been applied 
       to extrapolate the data along the coexistence curve. Circles show the 
       universal distribution of the 2D Ising universality class.
       ({\bf c})
       Same as ({\bf a}) but for film thickness $D=48$. The inverse critical temperature is $\epsilon_c=0.0625(10)$.
       ({\bf d})
       Same as ({\bf b}) but for film thickness $D=48$ and $\epsilon_c=0.0625$.
       }
    \end{minipage}%
\end{figure}

\begin{figure}[htbp]
    \begin{minipage}[b]{160mm}%
       \caption{
       \label{fig:REND} 
       ({\bf a}) Dependence of the effective interface potential on the lateral system size $L$ in
       a thin film of width $D=24$ at $\epsilon=0.065$
       ({\bf b}) same as ({\bf a}) but for $D=48$ and $\epsilon=0.069$. The inset presents an enlarged 
       view of the minimum close to the wall. The scale on the abscissa corresponds to the distance between the wall 
       and the interface in units of $R_g$.
       }
    \end{minipage}%
\end{figure}

\begin{figure}[htbp]
    \begin{minipage}[b]{160mm}%
       \caption{
       \label{fig:D24} 
       Free energy difference per unit area and $k_BT$ of the localized and delocalized
       state as a function of the lateral system size. The symbols represent the MC data, 
       while the solid lines are calculated from the effective interface Hamiltonian.
       The temperature was chosen such that $\Delta g \to 0$ for $L \to \infty$.
       }
    \end{minipage}%
\end{figure}

\begin{figure}[htbp]
    \begin{minipage}[b]{160mm}%
       \caption{
       \label{fig:TC} 
       Temperatures of the critical points and the triple point as a function of the film thickness $D$.
       Open symbols mark the results of the finite size scaling analysis. We have applied a correction 
       factor $(1+0.85/D)^{-1}$ to account for the film thickness dependence of the density at the center
       (filled symbols). Dashed lines are only guides to the eye. The arrow on the right hand side marks 
       the value of the wetting transition temperature
       obtained independently via the Young equation.\protect\cite{WET}
       }
    \end{minipage}%
\end{figure}

\begin{figure}[htbp]
    \begin{minipage}[b]{160mm}%
       \caption{
       \label{fig:PHASEN} 
       ({\bf a}) Phase diagram of a binary polymer blend ($N=32$). The upper curve shows the binodals
       in the infinite system; the middle one corresponds to a thin film of thickness $D=48$
       and symmetric boundary fields $\epsilon_w=0.16$, which both prefer species $A$. 
       The lower curve corresponds to a thin film with antisymmetric surfaces. The arrow marks
       the location of the wetting transition. Full circles mark critical points; open circles/dashed line
       denote the triple point.
       ({\bf b}) Coexistence curves in the ($T,\Delta \mu$)--plane. Circles mark critical points,
       and the diamond indicates the location of the wetting transition temperature. It is indistinguishable
       from the temperature of the triple point.
       }
    \end{minipage}%
\end{figure}


    \newpage
    \begin{minipage}[t]{160mm}%
       \mbox{
	({\bf a})
        \setlength{\epsfxsize}{8cm}
        \epsffile{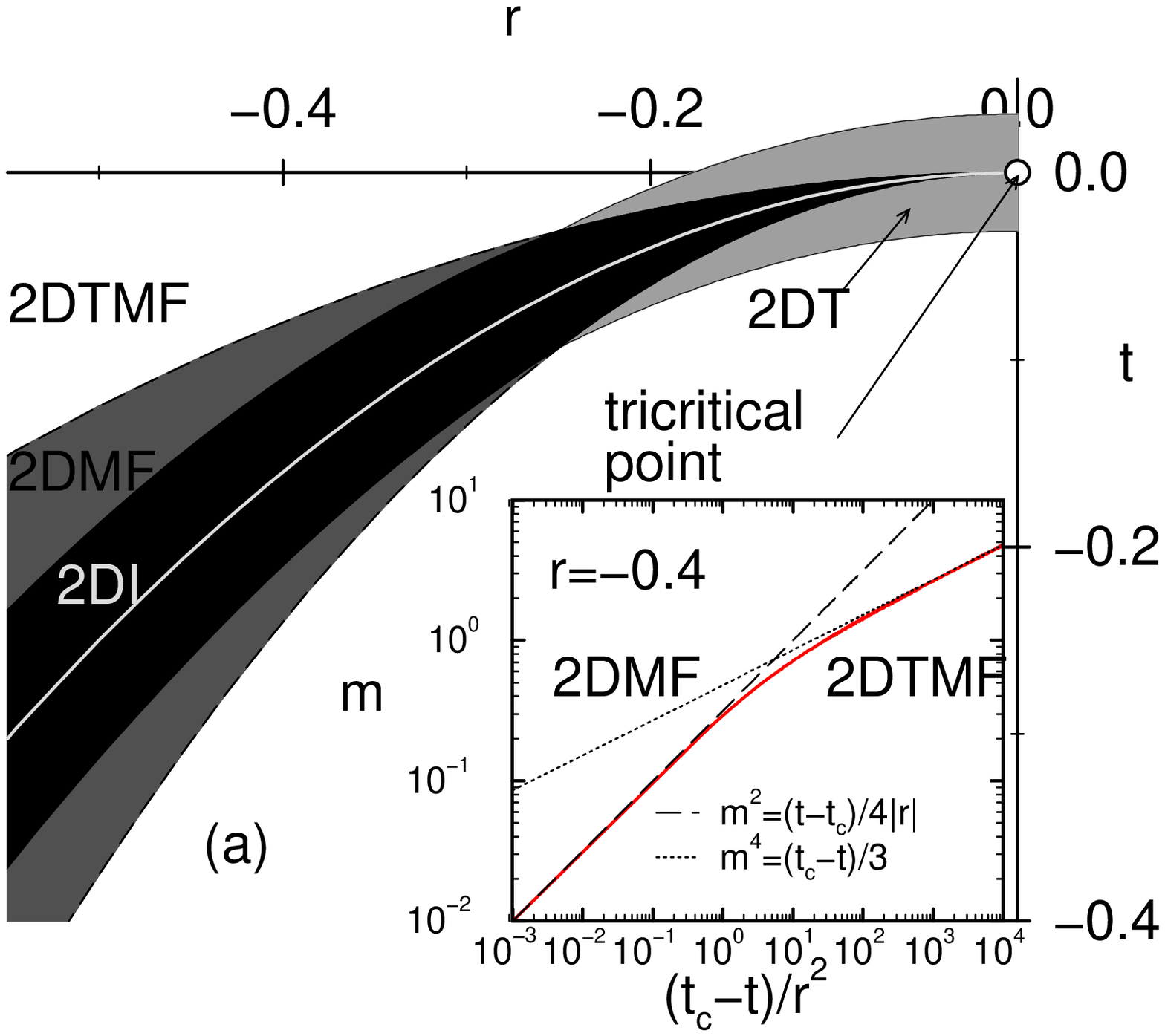}
	({\bf b})
        \setlength{\epsfxsize}{8cm}
        \epsffile{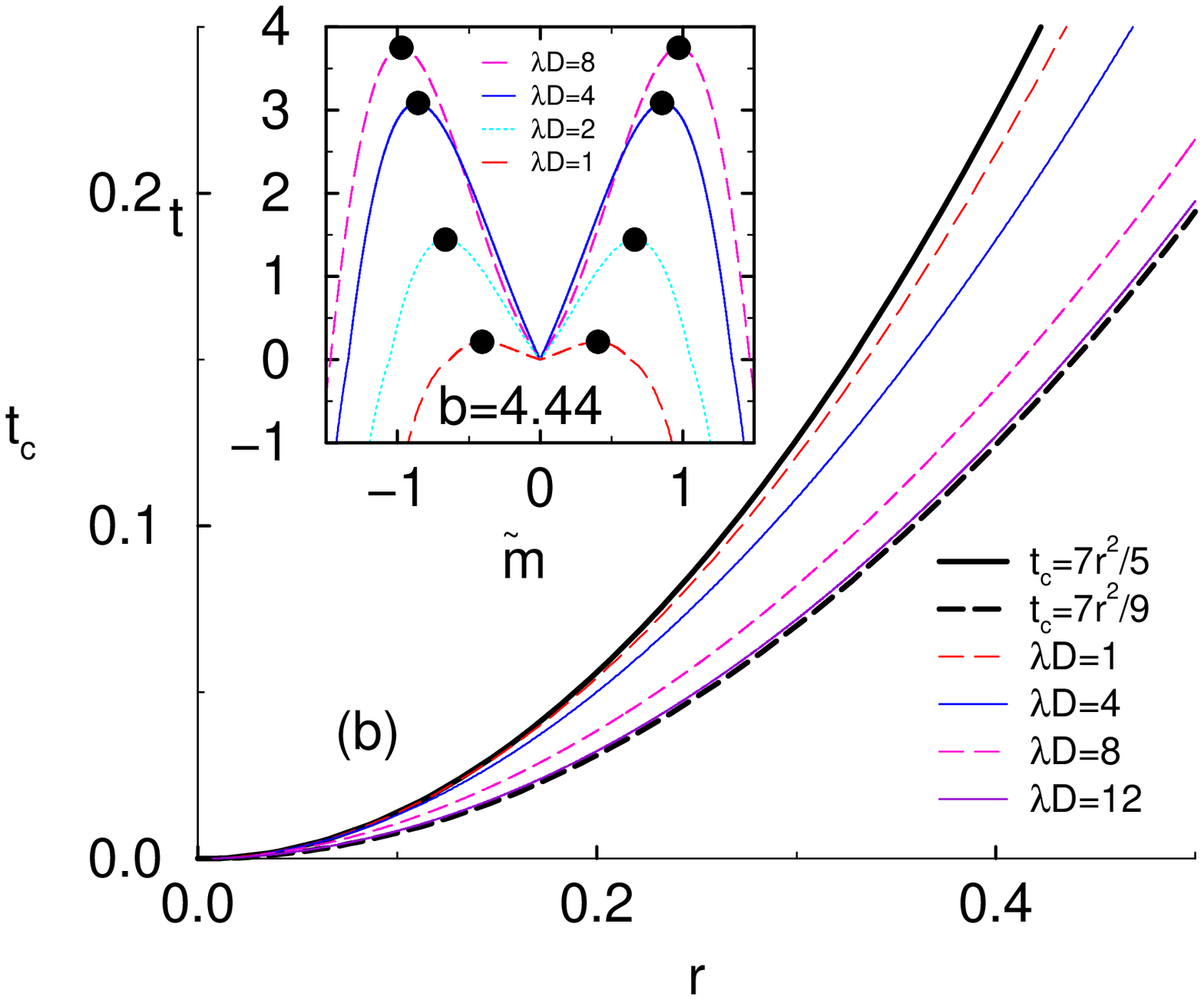}
       }
    \end{minipage}%

    \vspace*{1cm}
    {\tt Fig.1 : M.M{\"u}ller and K.Binder, Interface localisation--delocalisation in a symmetric ...}

    \newpage
    \begin{minipage}[t]{160mm}%
       \mbox{
        \setlength{\epsfxsize}{8cm}
        \epsffile{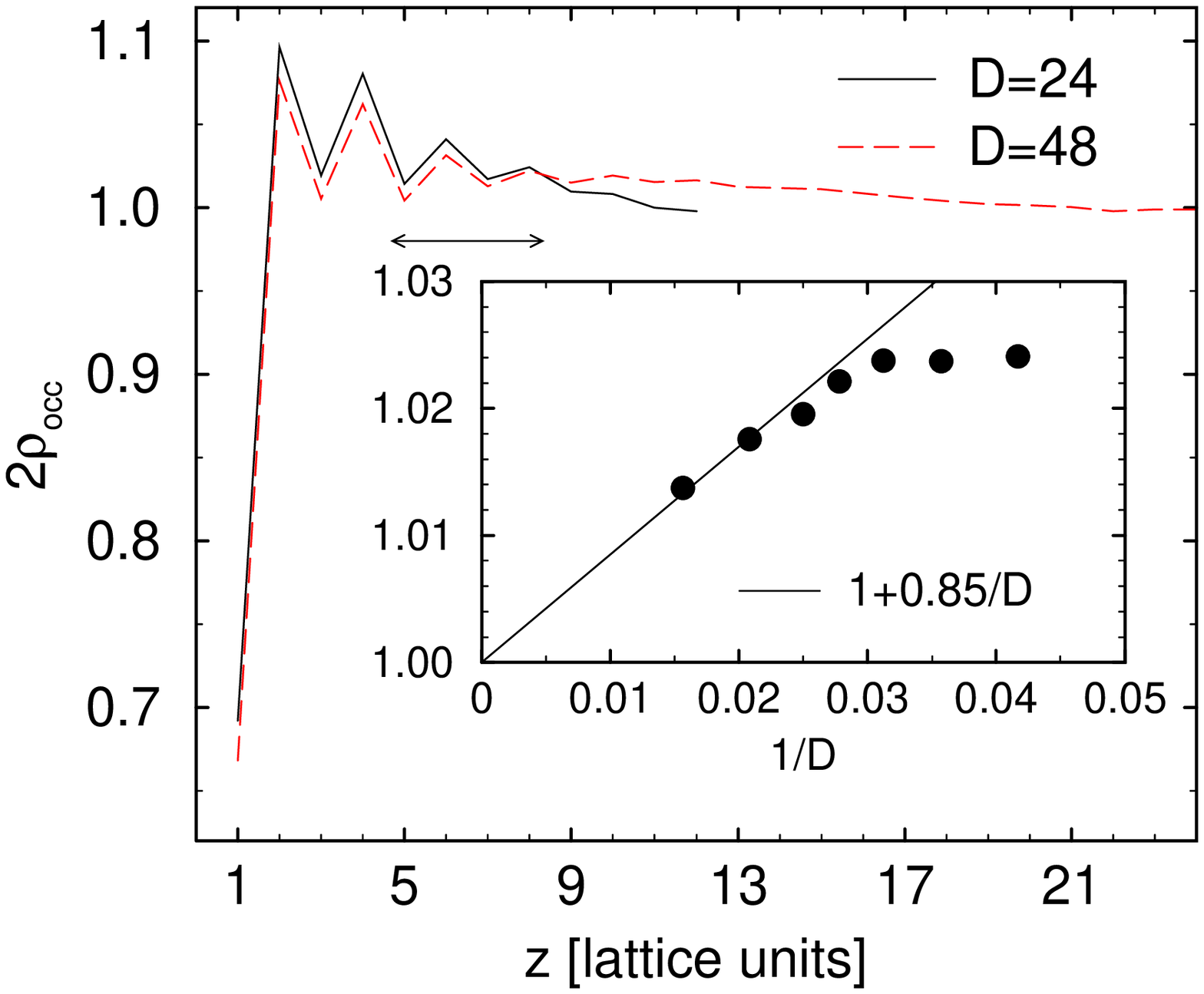}
       }
    \end{minipage}%

    \vspace*{1cm}
    {\tt Fig.2 : M.M{\"u}ller and K.Binder, Interface localisation--delocalisation in a symmetric ...}

    \newpage
    \begin{minipage}[t]{160mm}%
       \mbox{
	({\bf a})
        \setlength{\epsfxsize}{8cm}
        \epsffile{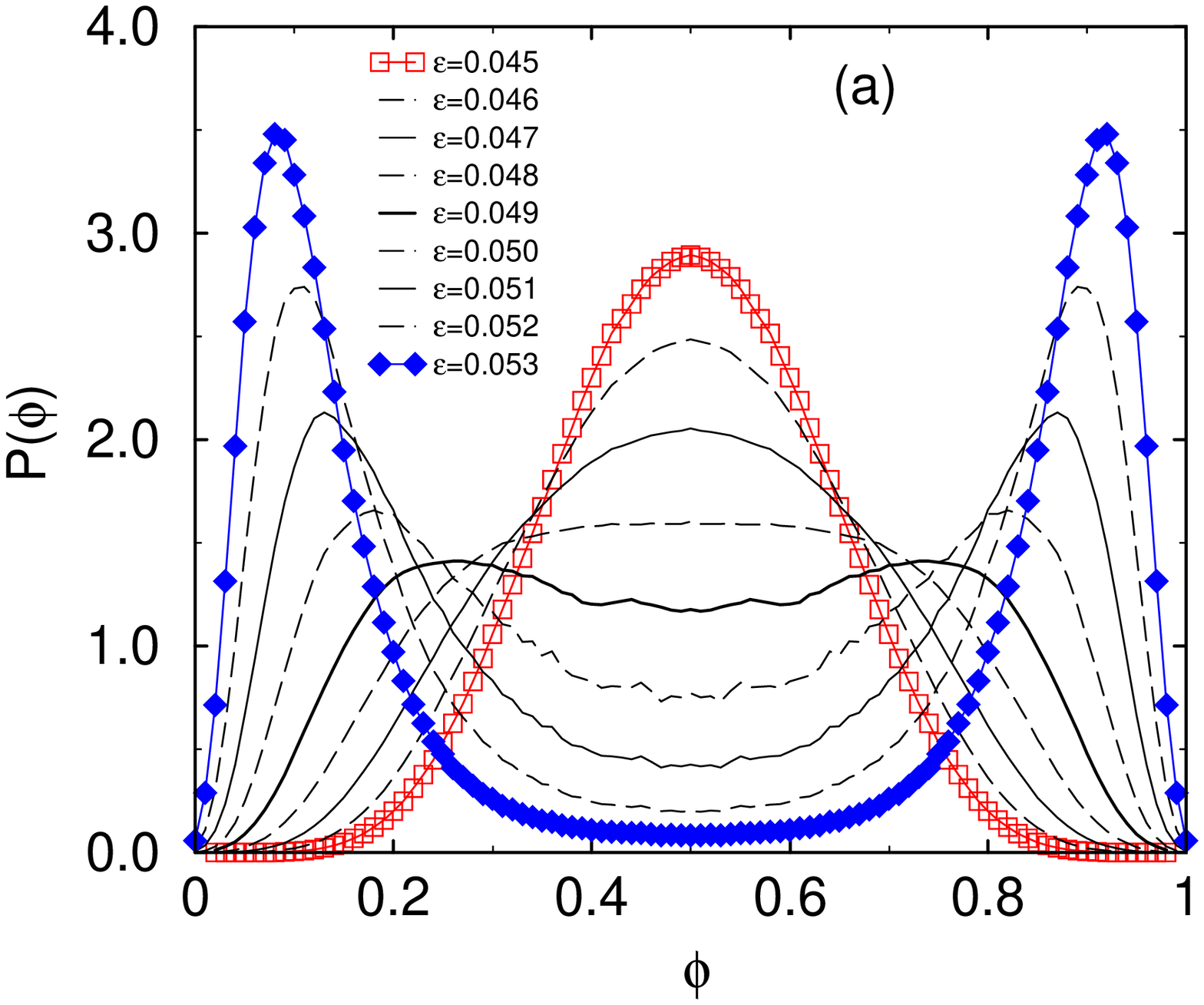}
	({\bf b})
        \setlength{\epsfxsize}{8cm}
        \epsffile{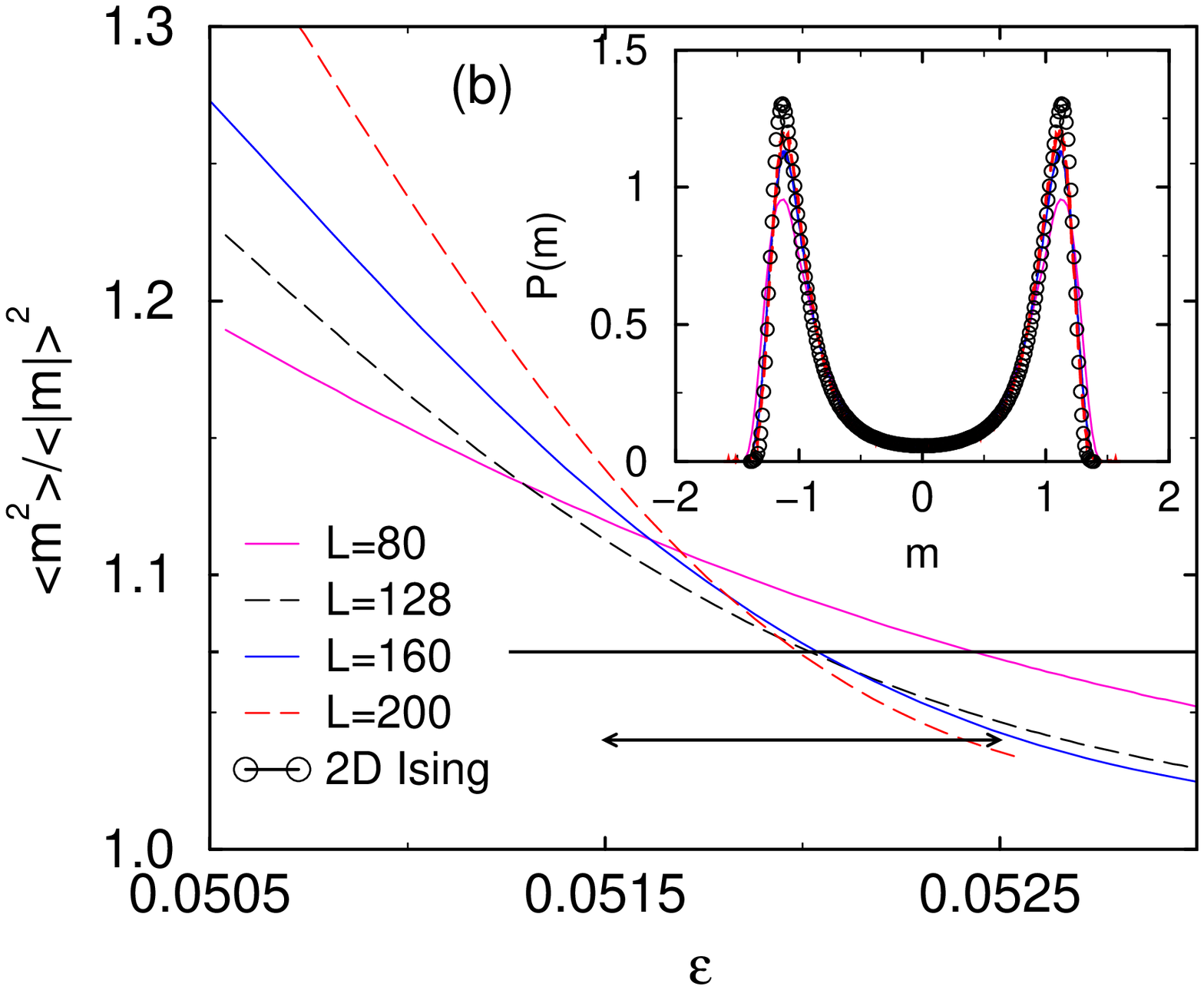}
       }\\
       \mbox{
	({\bf c})
        \setlength{\epsfxsize}{8cm}
        \epsffile{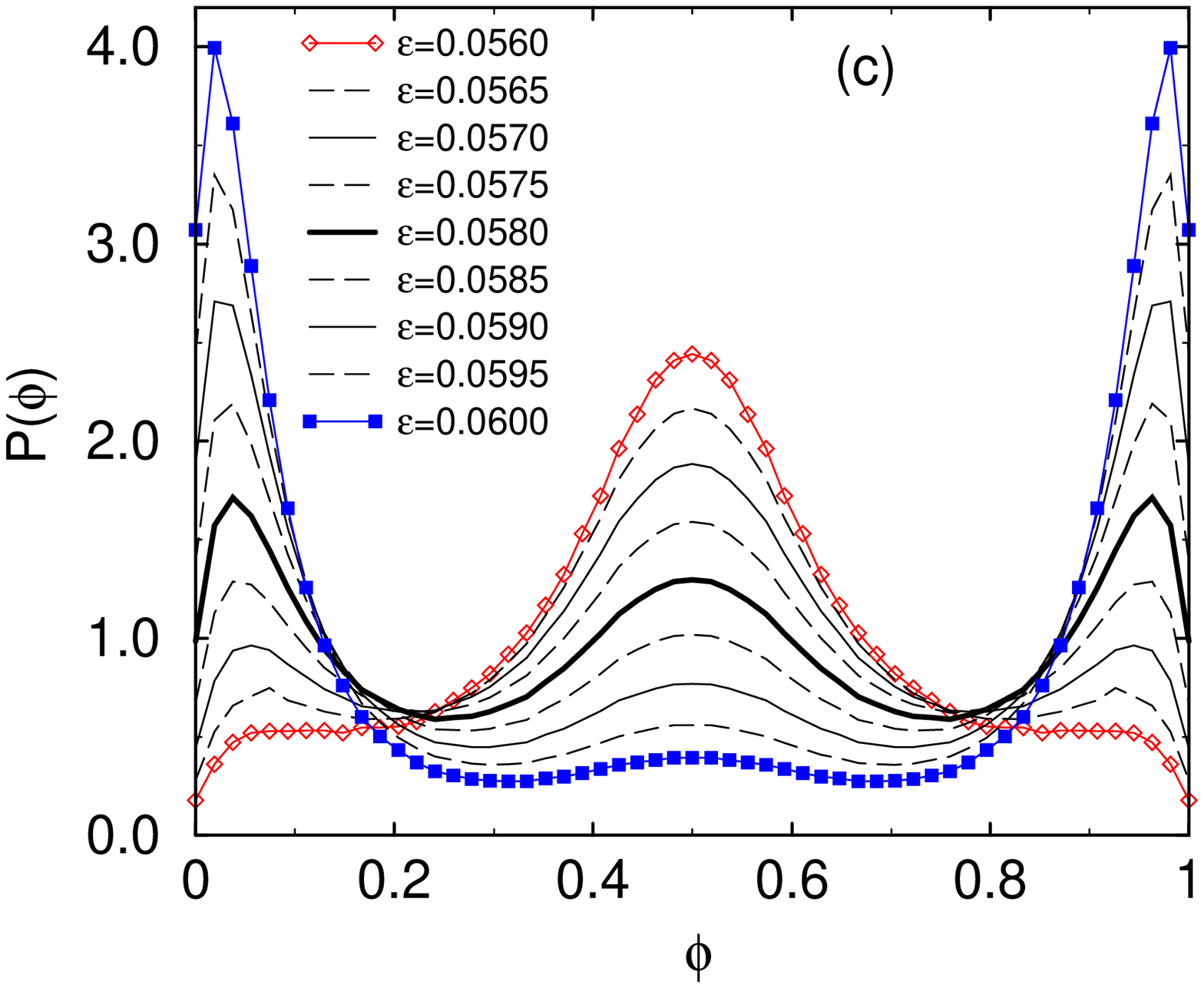}
	({\bf d})
        \setlength{\epsfxsize}{8cm}
        \epsffile{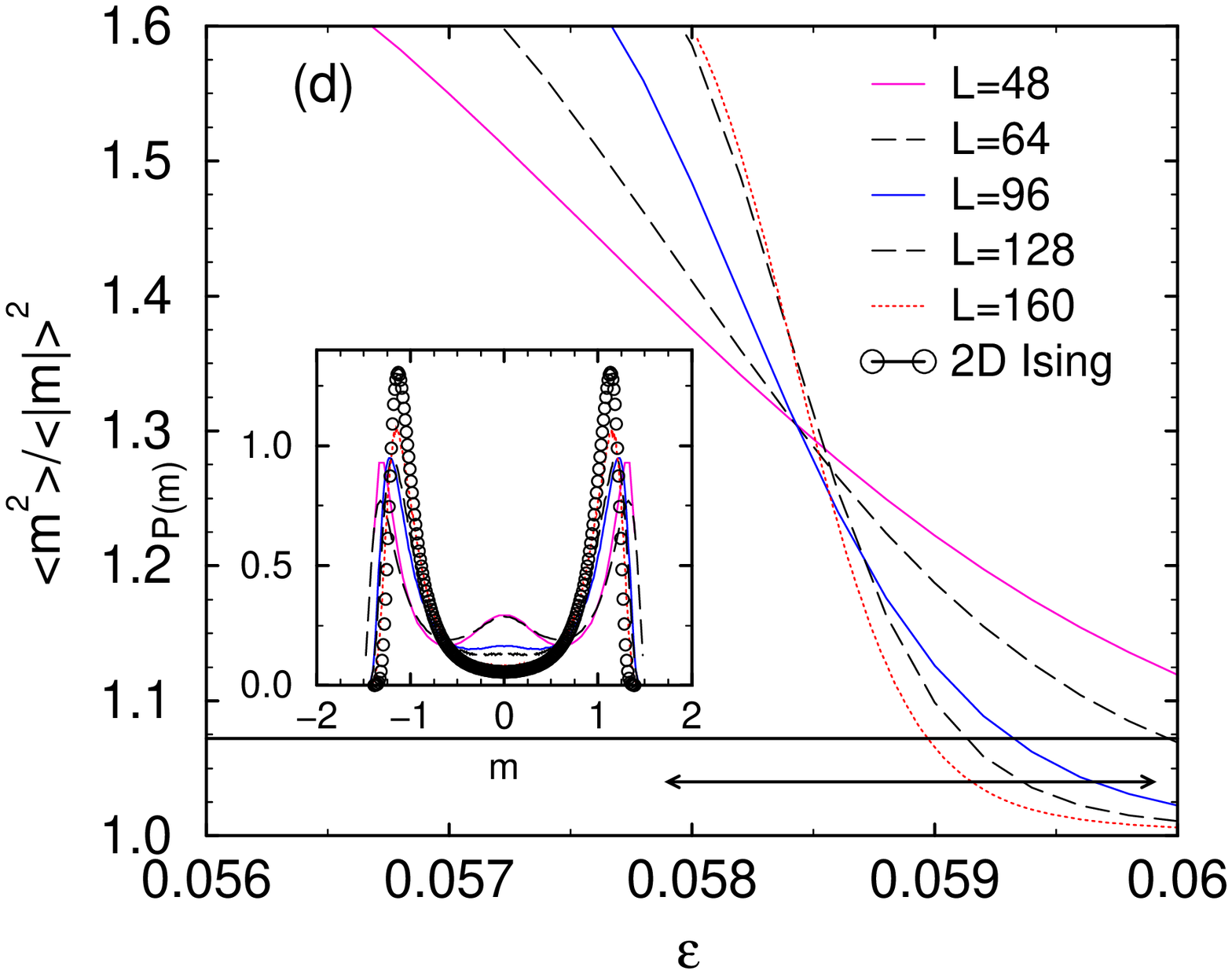}
       }
    \end{minipage}%

    \vspace*{1cm}
    {\tt Fig.3 : M.M{\"u}ller and K.Binder, Interface localisation--delocalisation in a symmetric  ...}

    \newpage
    \begin{minipage}[t]{160mm}%
       \mbox{
	({\bf a})
        \setlength{\epsfxsize}{8cm}
        \epsffile{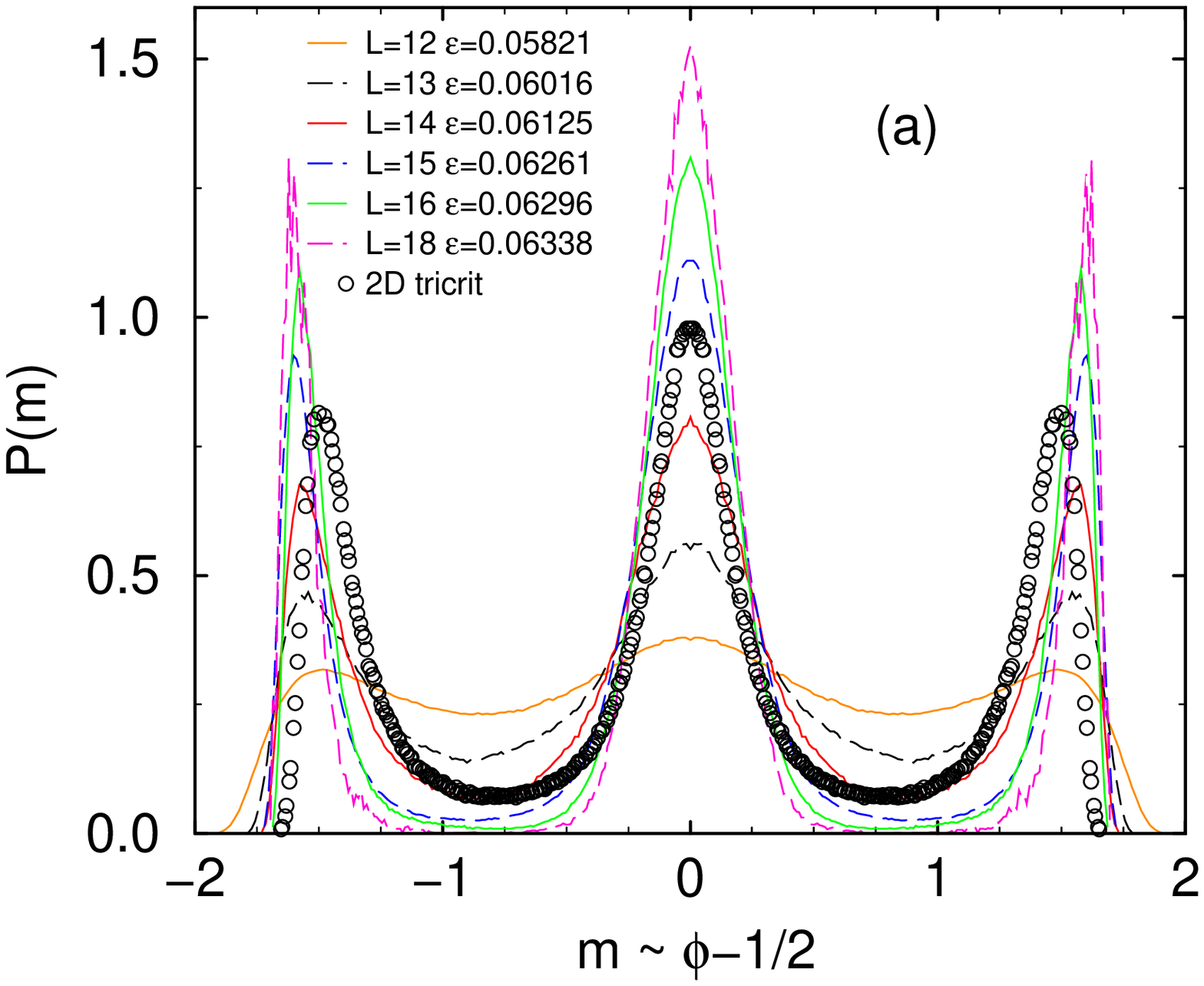}
	({\bf b})
        \setlength{\epsfxsize}{8cm}
        \epsffile{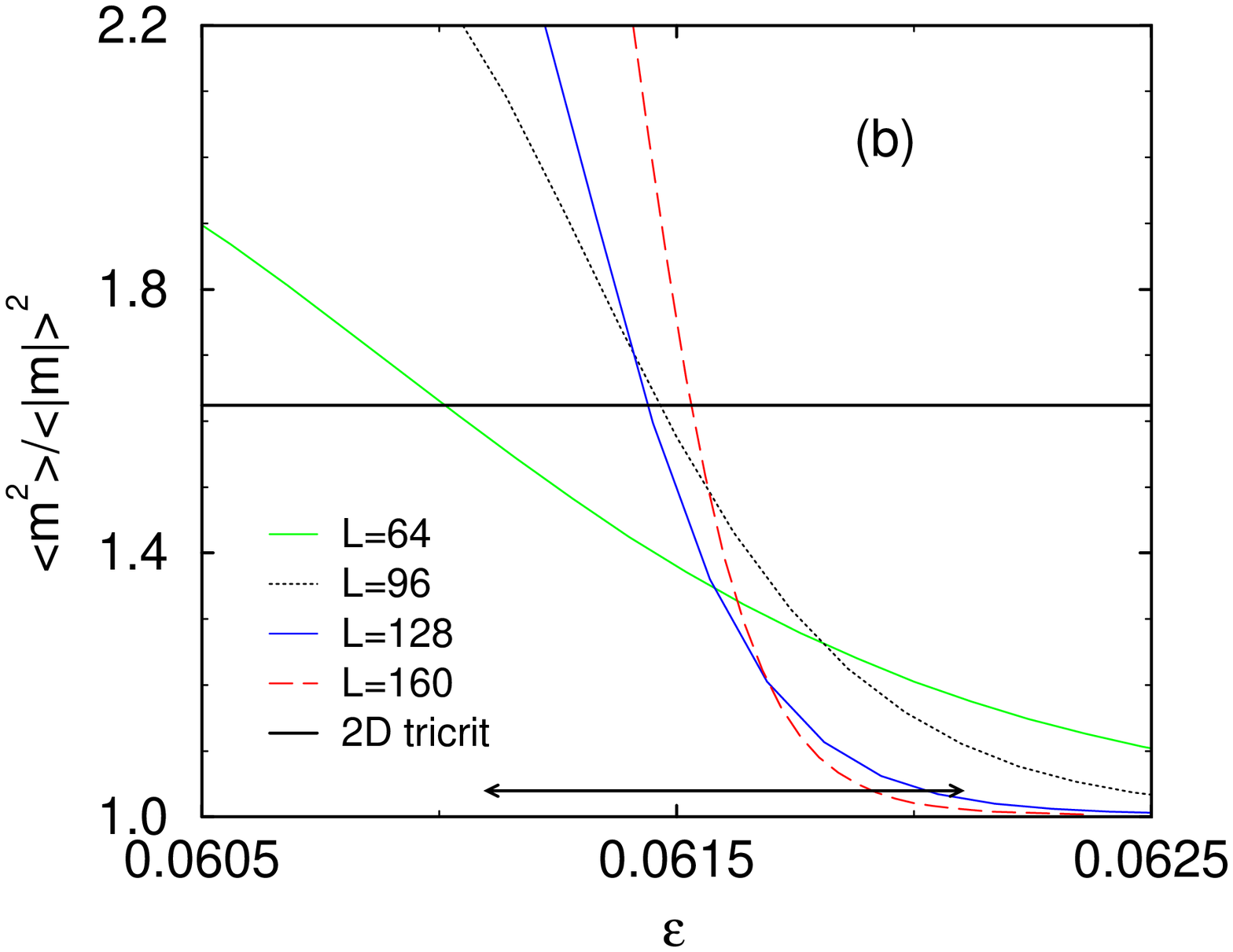}
       }\\
       \mbox{
	({\bf c})
        \setlength{\epsfxsize}{8cm}
        \epsffile{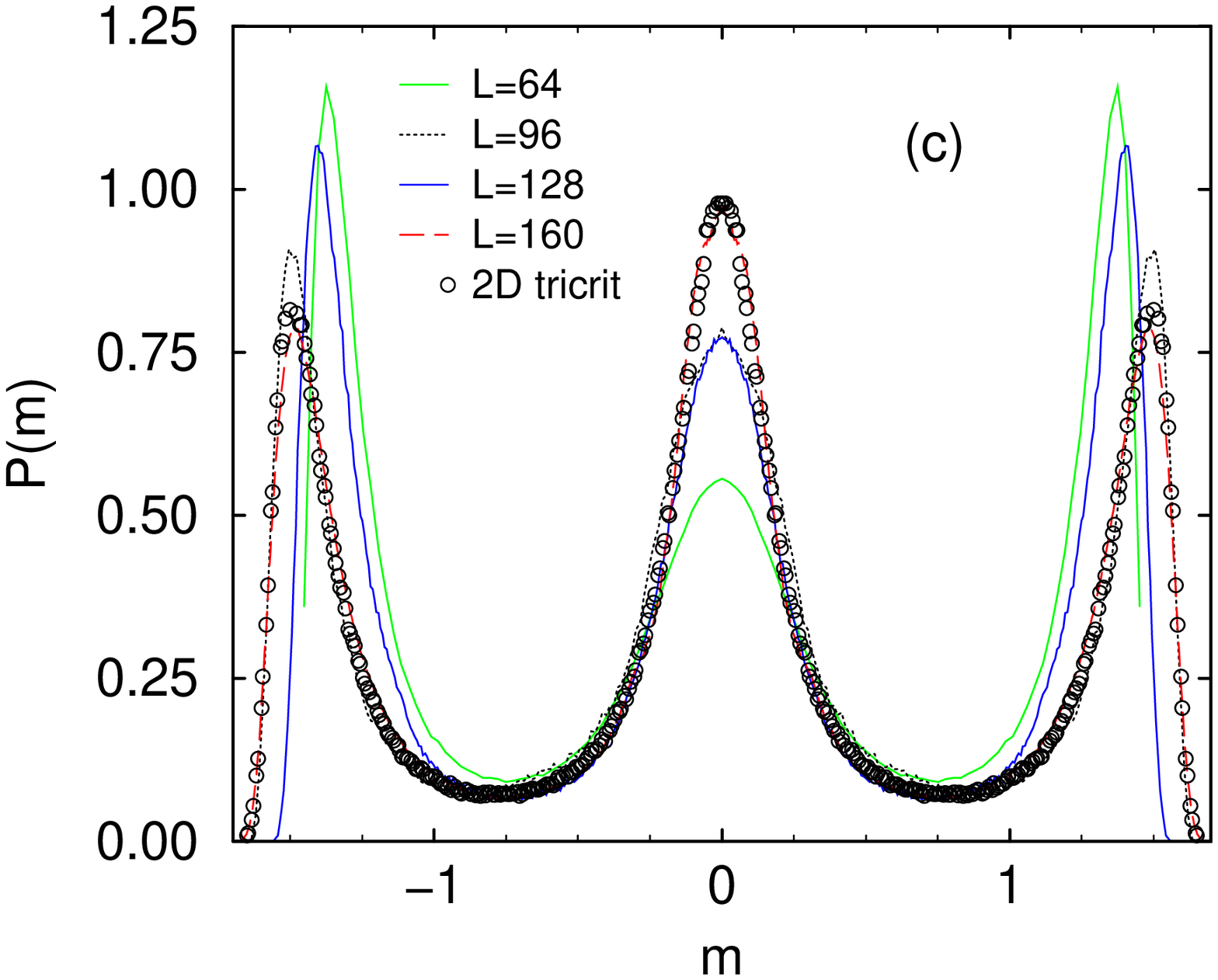}
        }
    \end{minipage}%

    \vspace*{1cm}
    {\tt Fig.4 : M.M{\"u}ller and K.Binder, Interface localisation--delocalisation in a symmetric  ...}

    \newpage
    \begin{minipage}[t]{160mm}%
       \mbox{
	({\bf a})
        \setlength{\epsfxsize}{8cm}
        \epsffile{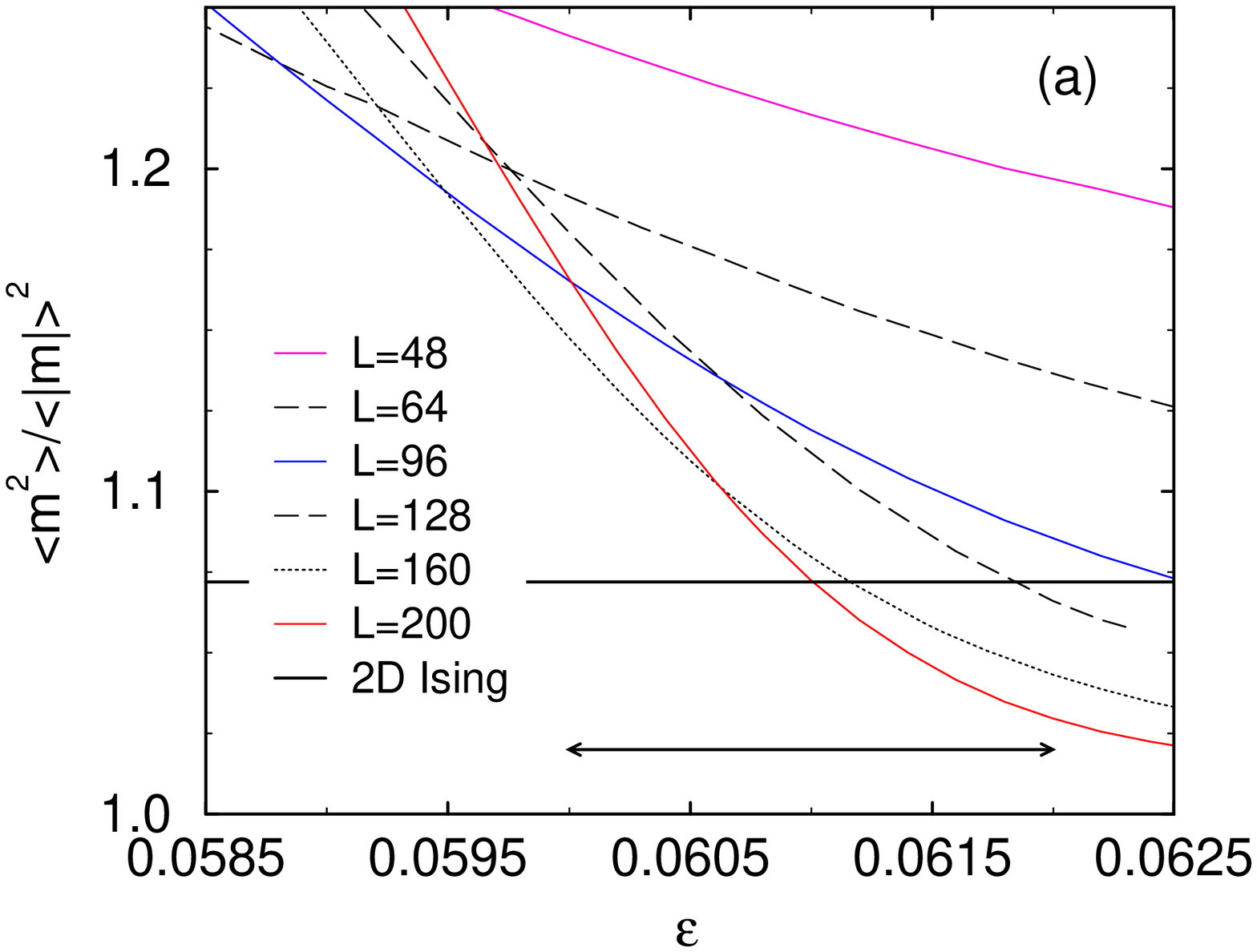}
	({\bf b})
        \setlength{\epsfxsize}{8cm}
        \epsffile{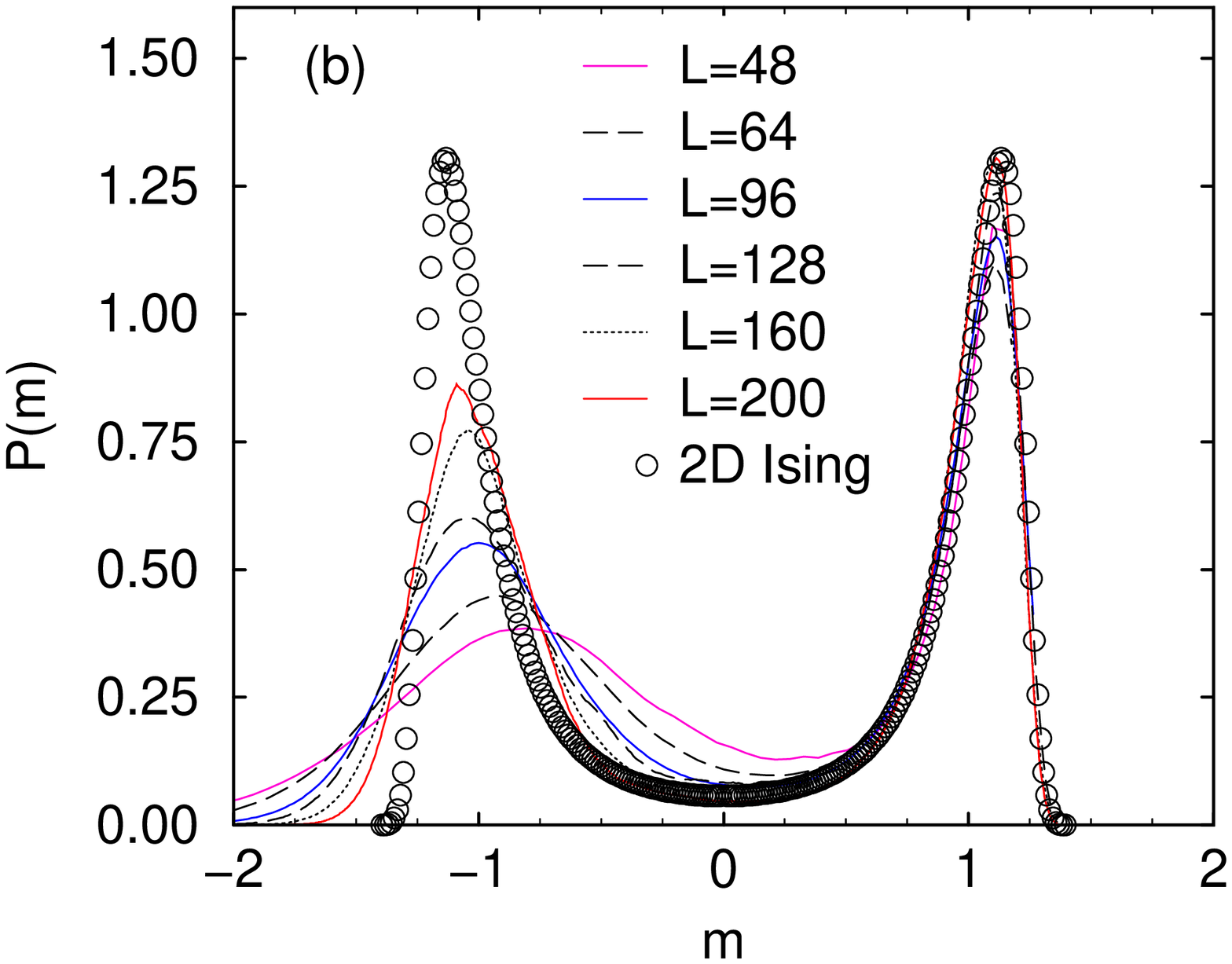}
       }
       \mbox{
	({\bf c})
        \setlength{\epsfxsize}{8cm}
        \epsffile{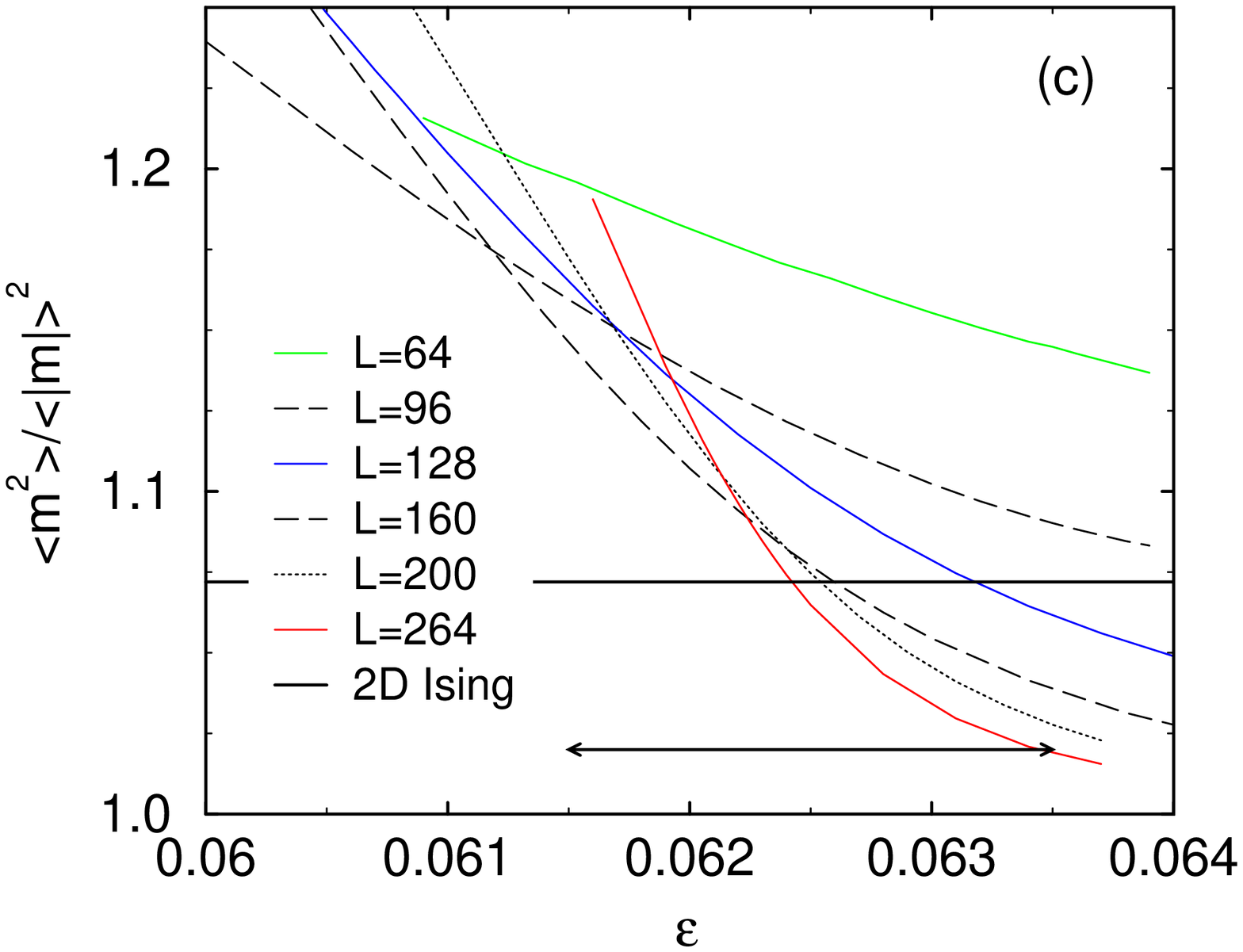}
	({\bf d})
        \setlength{\epsfxsize}{8cm}
        \epsffile{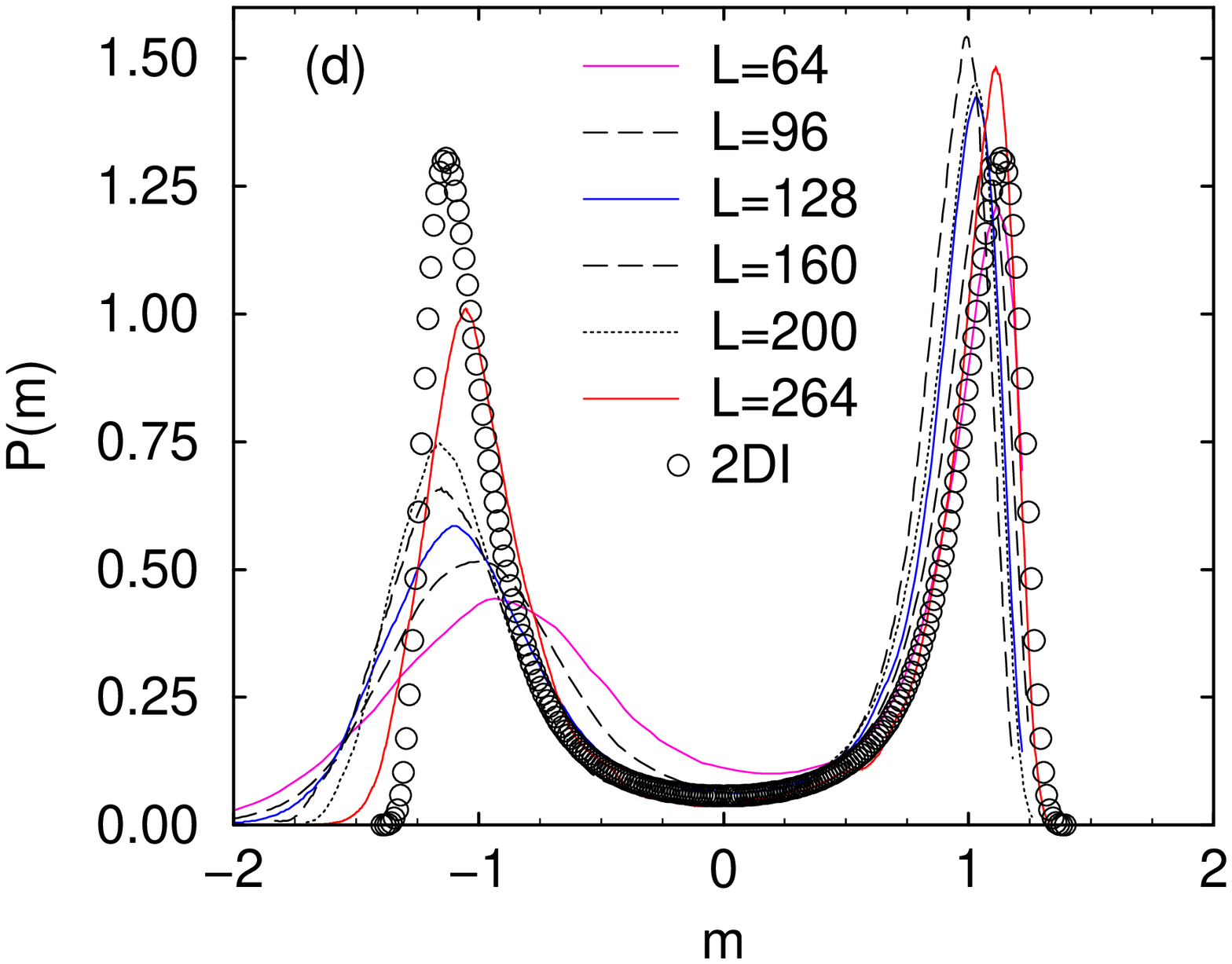}
       }
    \end{minipage}%

    \vspace*{1cm}
    {\tt Fig.5 : M.M{\"u}ller and K.Binder, Interface localisation--delocalisation in a symmetric  ...}

    \newpage
    \begin{minipage}[t]{160mm}%
       \mbox{
       ({\bf a})
        \setlength{\epsfxsize}{8cm}
        \epsffile{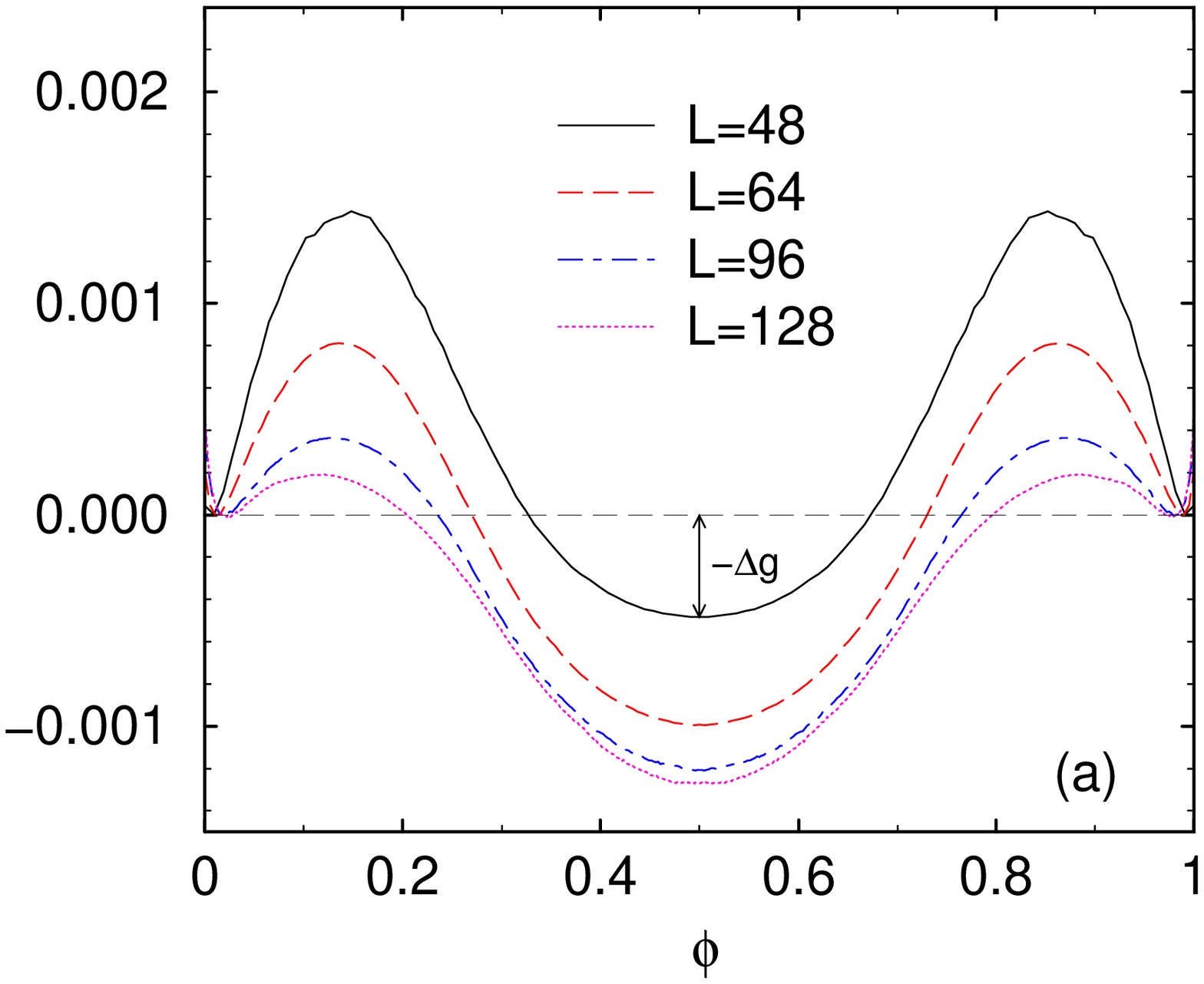}
       ({\bf b})
        \setlength{\epsfxsize}{8cm}
        \epsffile{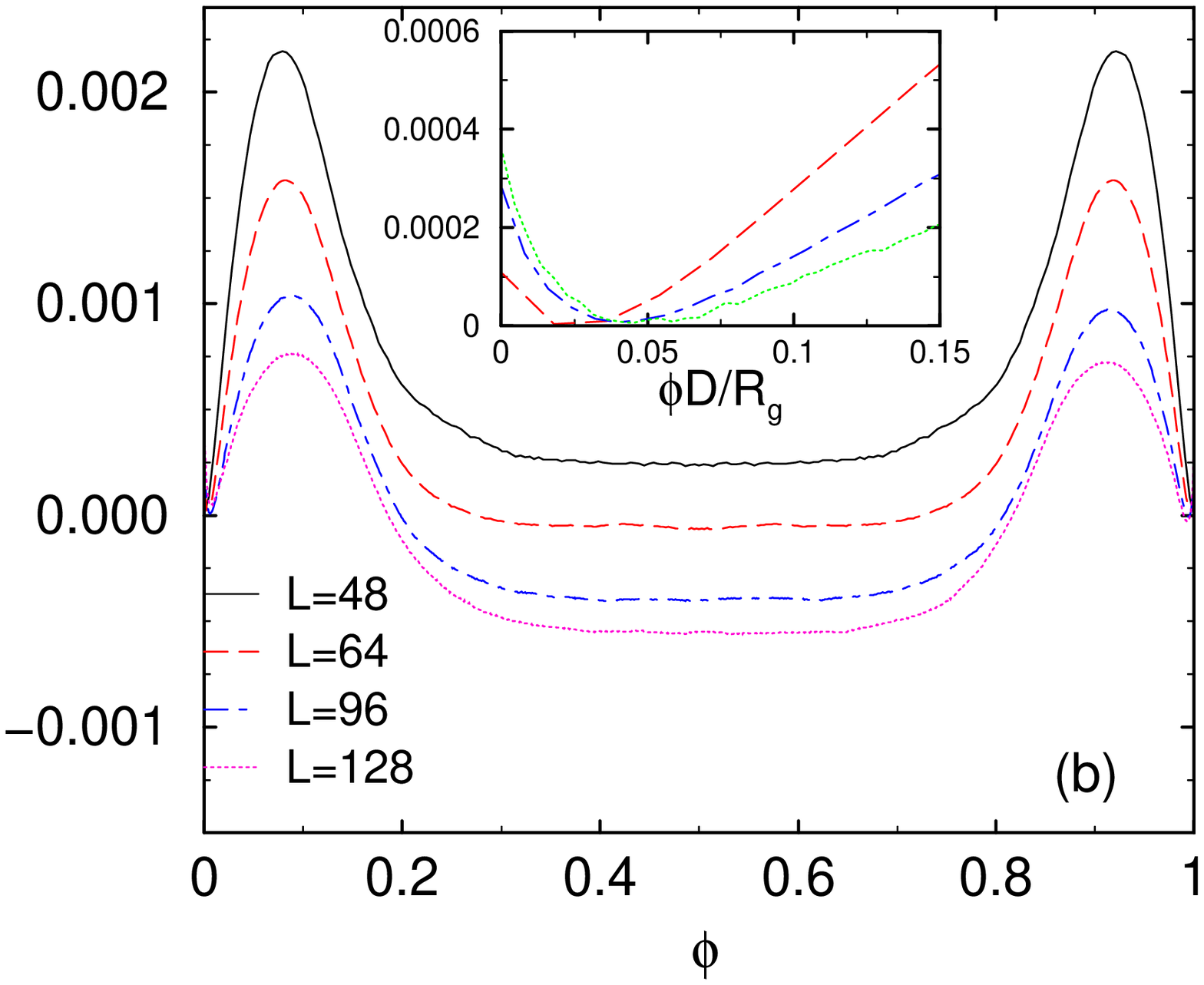}
       }
    \end{minipage}%

    \vspace*{1cm}
    {\tt Fig.6 : M.M{\"u}ller and K.Binder, Interface localisation--delocalisation in a symmetric  ...}

    \newpage
    \begin{minipage}[t]{160mm}%
       \mbox{
        \setlength{\epsfxsize}{8cm}
        \epsffile{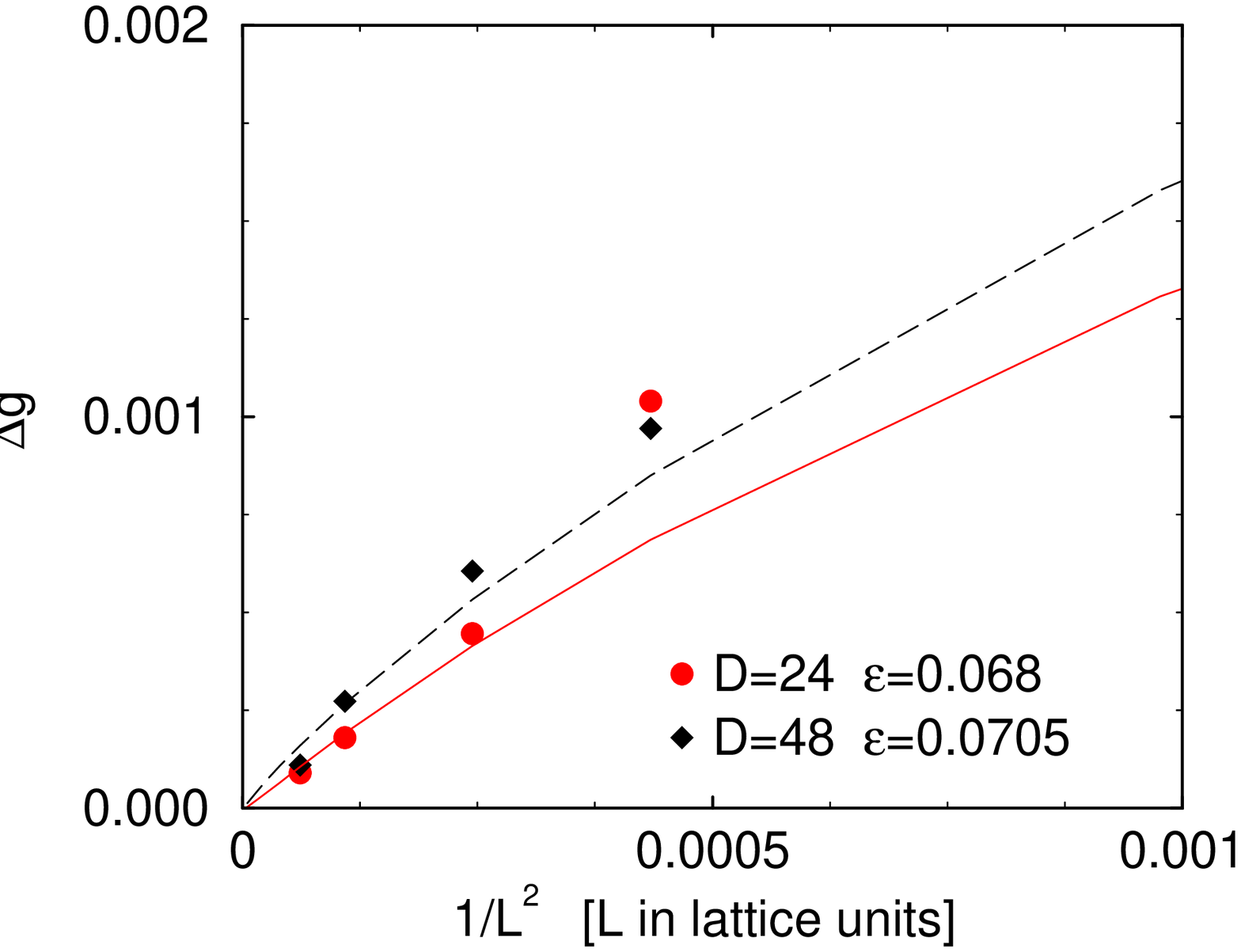}
       }
    \end{minipage}%

    \vspace*{1cm}
    {\tt Fig.7 : M.M{\"u}ller and K.Binder, Interface localisation--delocalisation in a symmetric  ...}

    \newpage
    \begin{minipage}[t]{160mm}%
       \mbox{
        \setlength{\epsfxsize}{8cm}
        \epsffile{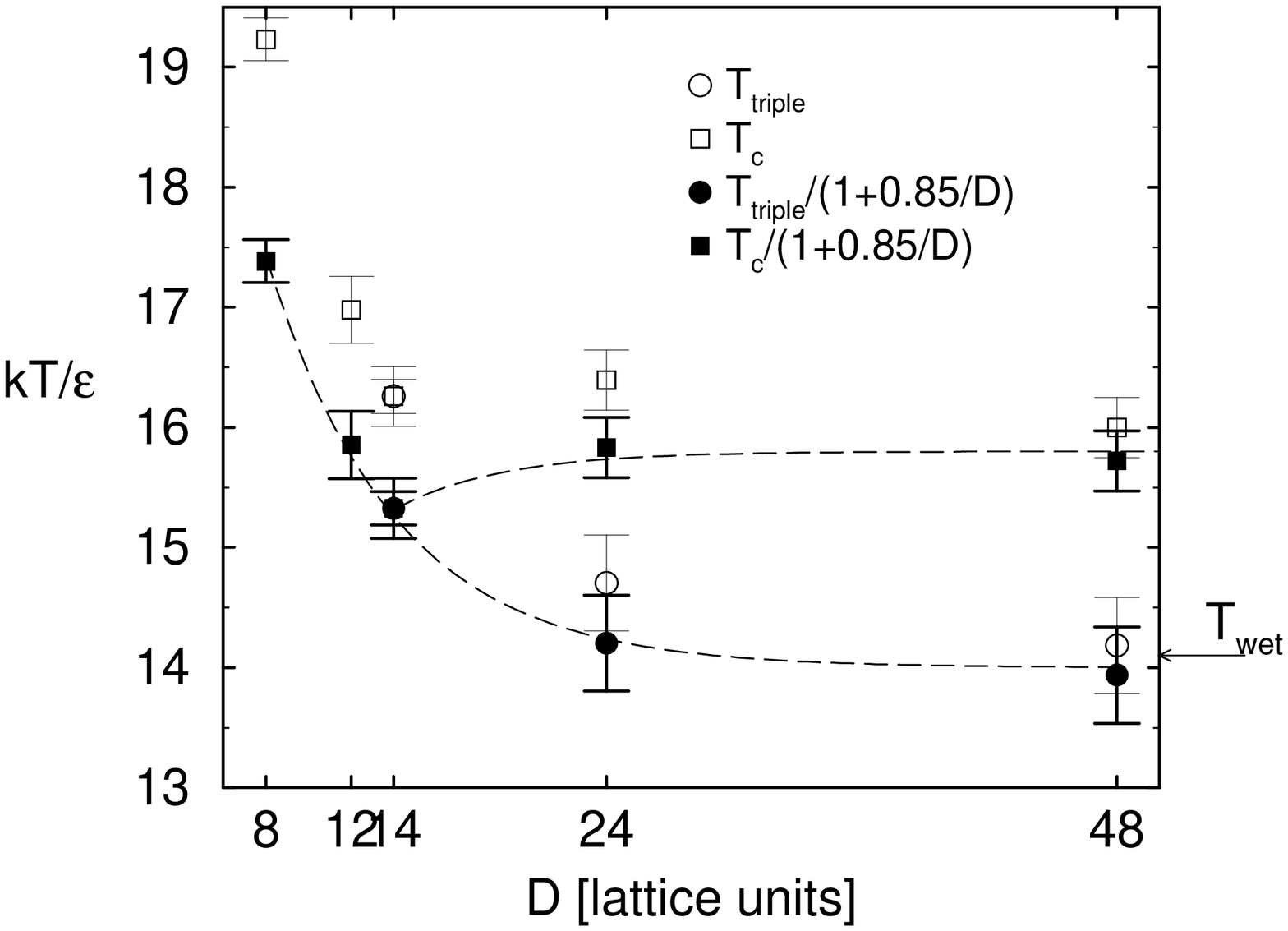}
       }
    \end{minipage}%

    \vspace*{1cm}
    {\tt Fig.8 : M.M{\"u}ller and K.Binder, Interface localisation--delocalisation in a symmetric  ...}

    \newpage
    \begin{minipage}[t]{160mm}%
       \mbox{
       ({\bf a})
        \setlength{\epsfxsize}{8.5cm}
        \epsffile{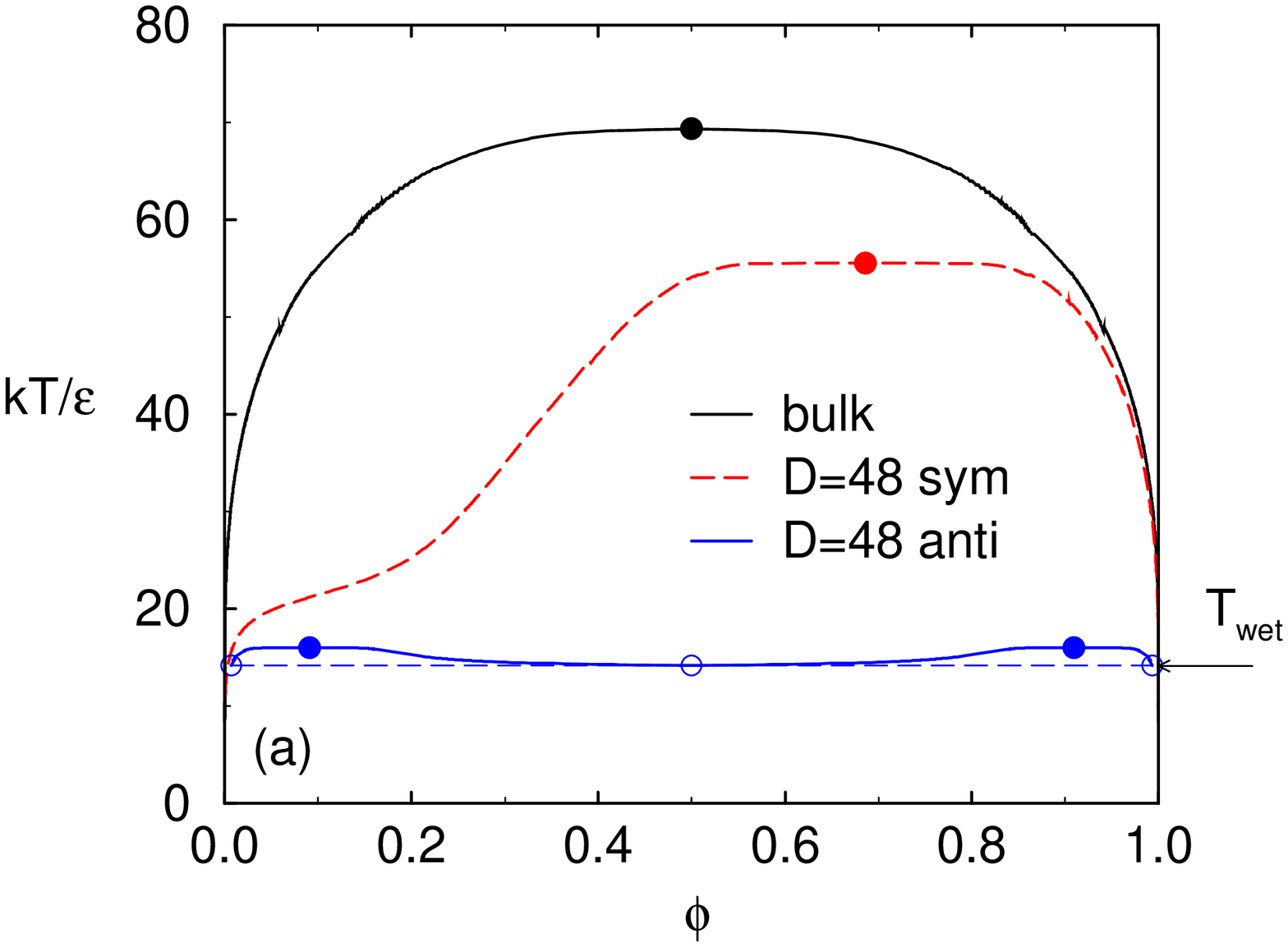}
       ({\bf b})
        \setlength{\epsfxsize}{8cm}
        \epsffile{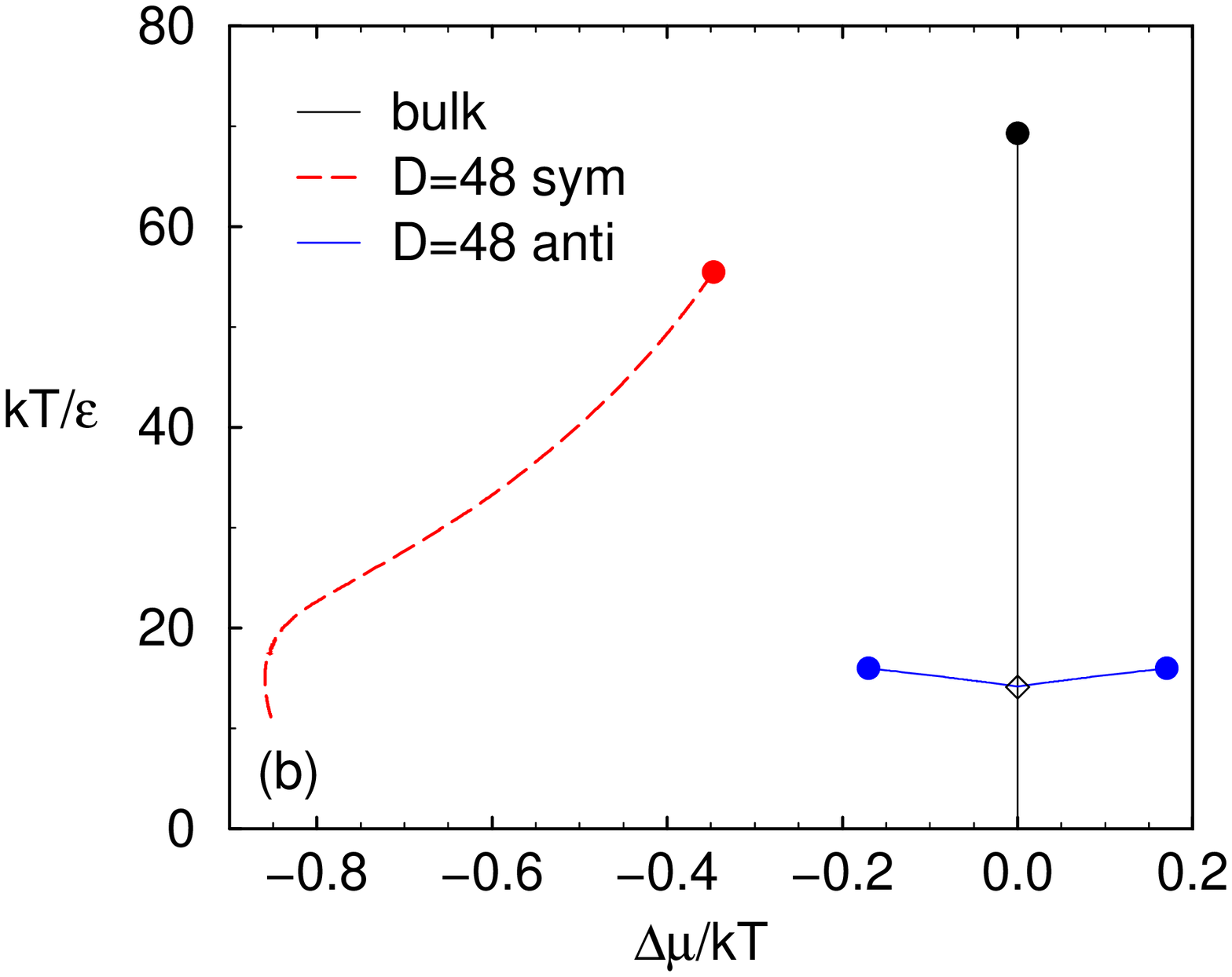}
       }
    \end{minipage}%

    \vspace*{1cm}
    {\tt Fig.9 : M.M{\"u}ller and K.Binder, Interface localisation--delocalisation in a symmetric polymer blend ...}


\newpage

\begin{thebibliography}{99}
\bibitem{EREV} R. Evans, J.Phys.Cond.Matter {\bf 2}, 8989 (1990).
\bibitem{PREV}  A.O. Parry, J.Phys.Condens.Matter {\bf 8}, 10761 (1996).
\bibitem{DIETRICH} S. Dietrich, Phase Transitions and Critical Phenomena, Vol 12, C. Domb and J. Lebowitz (eds) Academic Press, London (1988).
\bibitem{DEGENNES} P.G. De Gennes, Wetting statics and dynamics, Rev.Mod.Phys. {\bf 57}, 827 (1985).
\bibitem{POLYMER}  A. Budkowski, Adv. Polymer Sci. {\bf 148}, 1 (1999).  K. Binder, Adv. Polymer Sci. {\bf 138}, 1  (1999). 
\bibitem{NAKANISHI} M.E. Fisher and H. Nakanishi, J.Chem.Phys. {\bf 75}, 5857 (1981).
                    H. Nakanishi and M.E. Fisher, J.Chem.Phys. {\bf 78}, 3279 (1983).
\bibitem{BROCHARD} F. Brochard--Wyart and P.-G. de Gennes, Acad.Sci. Paris {\bf 297}, 223 (1983).
\bibitem{PE} A.O. Parry and R. Evans, Phys.Rev.Lett. {\bf 64}, 439 (1990), Physica {\bf A 181}, 250 (1992).
\bibitem{SWIFT} M.R. Swift, A.L. Owczarek, and J.O. Indekeu, Europhys.Lett. {\bf 14}, 475 (1991).
\bibitem{BINDER} K. Binder, D.P. Landau, and A.M. Ferrenberg, Phys.Rev.Lett. {\bf 74}, 298 (1995);
                 Phys.Rev. {\bf E 51}, 2823 (1995).
		 E.V. Albano, K. Binder, D.W. Heermann, and W. Paul, Surf.Sci. {\bf 233}, 151 (1989).
\bibitem{TROUBLE}   J.O. Indekeu, A.L. Owczarek, and M.R. Swift , Phys.Rev.Lett. {\bf 66}, 2174 (1991).
                    A.O. Parry and  R. Evans, Phys.Rev.Lett. {\bf 66}, 2175 (1991).
\bibitem{GRAVITY}   J. Rogiers and J.O. Indekeu, Europhys.Lett. {\bf 24}, 21 (1993).
                    E. Carlon and A. Drzewinski, Phys.Rev.Lett. {\bf 79}, 1591 (1997).
\bibitem{NEW}  K. Binder, R. Evans, D.P. Landau, and A.M. Ferrenberg, Phys.Rev. {\bf E 53}, 5023 (1996).
\bibitem{ANDREAS} A. Werner, F. Schmid, M. M\"uller, and K. Binder, J.Chem.Phys. {\bf 107}, 8175 (1997).
\bibitem{KLEIN} T. Kerle, J. Klein, and K. Binder, Phys.Rev.Lett. {\bf 77}, 1318 (1996), Europ. Phys. J. B{\bf 7},  401 (1999).
\bibitem{SF}    M. Sferrazza, M. Heppenstall--Butler, R. Cubitt, D. Bucknall, J. Webster, R.A.L. Jones, Phys.Rev.Lett. {\bf 81}, 5173 (1998);
                  M. Sferrazza, C. Xiao, R.A.L. Jones, G.D. Bucknall, J. Webster, J. Penfold, Phys.Rev.Lett. {\bf 78}, 3693 (1997).
\bibitem{MSCF1}   M. M\"uller, K. Binder, and E.V. Albano, Europhys.Lett. {\bf 49}, 724 (2000);
\bibitem{MSCF2}   M. M\"uller, K. Binder, and E.V. Albano, Physica {\bf A 279}, 188 (2000);
                  M. M\"uller, E.V. Albano, and K. Binder, Phys.Rev. {\bf E} (in press).
\bibitem{B2} A.M. Ferrenberg, D.P. Landau, and K. Binder, Phys.Rev. {\bf E 58}, 3353 (1998).
\bibitem{MREV}  M. M\"uller, Macromol. Theory Simul. (Feature Article) {\bf 8}, 343 (1999).
\bibitem{WET}   M. M\"uller and K. Binder, Macromolecules {\bf 31}, 8323 (1998).
\bibitem{TRI1}  B. Nienhuis, A.N. Nerker, E.K. Riedel, and M. Schick, Phys.Rev.Lett. {\bf 43}, 737 (1979).
\bibitem{TRI2}  M.P.M. den Nijs, J.Phys. {\bf A 12}, 1857 (1979).
\bibitem{TRI3}  R.B. Rearson, Phys.Rev. {\bf B 22}, 2579 (1980).
\bibitem{GINZBURG}   V.L. Ginzburg, Sov.Phys.Solid State  {\bf 1}, 1824 (1960);
                     P.G. de Gennes,  J.Phys.Lett. (Paris) {\bf 38}, L-441 (1977);
		     J.F. Joanny,  J.Phys.A  {\bf 11}, L-117 (1978);
		     K. Binder,  Phys.Rev.A  {\bf 29}, 341 (1984).
\bibitem{BFM}     I. Carmesin and K. Kremer, Macromolecules {\bf 21}, 2819 (1988).
\bibitem{HPD} H.-P. Deutsch and K. Binder,  J. Chem. Phys. {\bf 94}, 2294 (1991).
\bibitem{MAP} J. Baschnagel {\em et al}, Adv.Polym.Sci. {\bf 152}, 41 (2000).
              V. Tries, W. Paul, J. Baschnagel, and K. Binder,J.Chem.Phys. {\bf 106}, 738 (1997).
\bibitem{MBO} M. M\"uller, K. Binder, and W.Oed, J.Chem.Soc. Faraday Trans. 91, 2369 (1995).
\bibitem{M0}  M. M\"uller and K. Binder, Macromolecules {\bf 28}, 1825 (1995). 
\bibitem{SARIBAN} A. Sariban and K. Binder, J. Chem. Phys. {\bf  86}, 5859 (1987).
\bibitem{REWEIGHT} B.A. Berg and T. Neuhaus, Phys.Rev.Lett {\bf 68}, 9 (1992);
              J. Lee, Phys.Rev.Lett {\bf 71}, 221 (1993);
	      E. Marinari and G. Parisi, Europhys.Lett. {\bf 19}, 451 (1992).
\bibitem{HISTO} A.M. Ferrenberg and R.H. Swendsen, Phys. Rev. Lett. {\bf  61},  2635 (1988);. ibid {\bf 63}, 1195 (1989);
              A. Bennett, J.Comp.Phys {\bf 22} 245 (1972).
\bibitem{BRUCEWILDING} A.D. Bruce and N.B. Wilding, Phys.Rev.Lett. {\bf 68}, 193 (1992);
                       N.B. Wilding, Phys.Rev. {\bf E 52}, 602 (1995).
\bibitem{FSS} K. Binder, Z.Phys {\bf B 43}, 119 (1981); Phys.Rev.Lett {\bf  47}, 693 (1981); Rep.Prog.Phys. {\bf 60}, 487 (1997).
\bibitem{NIGEL} N.B. Wilding and P. Nielaba, Phys. Rev. {\bf E 53}, 926 (1996).  
\bibitem{BC} M. Blume, Phys.Rev. {\bf 141}, 517 (1966); H.W. Capel, Physica {\bf 32}, 966 (1966).
\bibitem{EWR} C. Borgs and K. Kotecky, J.Stat.Phys. {\bf 61}, 79 (1990); Phys.Rev.Lett. {\bf 68}, 1734 (1992).
\bibitem{HPD2} H.-P. Deutsch and K. Binder, Macromolecules {\bf 25}, 6214 (1992).
\bibitem{YAN} Y. Rouault, J. Baschnagel, and K. Binder, J.Stat.Phys. {\bf 80}, 1009 (1995).
\bibitem{BWE}  A.D. Bruce and N.B. Wilding, Phys.Rev. {\bf E 60}, 3748 (1999). 
\bibitem{CAHN} J.W. Cahn, J.Chem.Phys. {\bf 66}, 3667 (1977). 
\bibitem{MICHAEL} M. Schick, Liquids at Interfaces, Les Houches, Session XLVIII, J. Charvolin, J.F. Joanny, and J. Zinn-Justin (eds), Elsevier, Amsterdam (1990).
\bibitem{LIPOWSKI} R. Lipowski, D.M. Kroll, and R.K.P. Zia, Phys.Rev. {\bf B 27}, 499 (1983).
\bibitem{BREZIN} E. Brezin, B.I. Halperin, and S. Leibler, Phys.Rev.Lett. {\bf 50}, 1387 (1983).
\bibitem{OMEGA1} D.S. Fisher and D.A. Huse, Phys.Rev. {\bf B 32}, 247 (1985).
\bibitem{OMEGA2} M.E. Fisher and H. Wen, Phys.Rev.Lett. {\bf 68}, 3654 (1992).
\bibitem{IH} D.A. Huse, W. van Saarloos, and J.D. Weeks, Phys.Rev. {\bf B 32}, 233 (1985).
\bibitem{HS} E.H. Hauge and M. Schick, Phys.Rev. {\bf B 27}, 4788 (1983).
\bibitem{TRIPLE1} D. Nicolaides and R. Evans, Phys.Rev. {\bf B 39}, 9336 (1989).
                  R. Evans and U. Marini Bettolo Marconi, Phys. Rev. {\bf A 32}, 3817 (1985).
\bibitem{EXP1} W.Zhao, M.H. Rafailovich, J. Sokolov, L.J. Fetters, R. Plano, M.K. Sanyal, and S.K. Sinha, Phys.Rev.Lett. {\bf 70}, 1453 (1993).
\bibitem{EXP3} J. Rysz, A. Budkowski, A. Bernasik, J. Klein, K. Kowalski, J. Jedlinski, and L.J. Fetters, Europhys.Lett. {\bf 50}, 35 (2000).
\bibitem{EXP2} M. Geoghegan, H. Ermer, G. J{\"u}ngst, G. Krausch, and R. Brenn , Phys.Rev. {\bf E 62}, 940 (2000).
\end{thebibliography}
\end{document}